\documentclass[useAMS,usenatbib,fleqn]{mnras}
\usepackage{amsmath}
\usepackage{graphicx}
\usepackage{epstopdf}
\usepackage{amssymb}
\usepackage{xcolor}
\usepackage{extarrows}
\usepackage{color}
\usepackage{amsmath,bm}
\usepackage{CJK}
\usepackage{hyperref}
\usepackage{forest}
\usepackage{tabularx}
\usepackage{ulem}
\newcommand{\pc}{pc}
\newcommand{\kpc}{kpc}
\newcommand{\Mpc}{Mpc}

\newcommand{\bh}{\mathrm{bh}}
\newcommand{\msun}{\mathrm{M_\odot}}

\newcommand{\imbh}{\mathrm{IMBH}}
\newcommand{\cm}{\mathrm{cm}}

\newcommand{\yw}[1]{#1}
\newcommand{\comment}[1]{}

\def\e1{e_1^2}

\begin{document}

\title[Hypervelocity binaries]{Hypervelocity binaries from close encounters with a SMBH-IMBH binary: orbital properties and diagnostics}
\author[Wang et al.]
{
Yi-Han Wang$^{1}$\thanks{E-mail:rosalba.perna@stonybrook.edu;\newline\qquad yihan.wang.1@stonybrook.edu},
Nathan Leigh$^{1}$, Alberto Sesana$^{2}$, Rosalba Perna$^{1}$
\\
$^{1}$ Department of Physics and Astronomy, Stony Brook University, Stony Brook, NY 11794-3800, USA\\
$^{2}$ School of Physics and Astronomy and Institute of Gravitational Wave Astronomy, University of Birmingham, Edgbaston B15 2TT, UK
}

\pagerange{\pageref{firstpage}--\pageref{lastpage}} \pubyear{2018}

\maketitle
   
\label{firstpage}

\begin{abstract}
  {Hypervelocity binaries (HVBs) can be produced by a close encounter between stellar binaries and supermassive black hole (SMBH)-intermediate massive black hole (IMBH) binaries. We perform scattering experiments between tight stellar binaries and an SMBH-IMBH binary (within the observationally allowed parameters), using high-precision N-body integrations. Simulation results show that only a few HVBs can be produced in the Galactic center if it harbors a $4\times 10^{3}$M$_\odot$ IMBH. However, we find that tight HVBs can be efficiently produced in prograde coplanar orbits with respect to the SMBH-IMBH binary. Therefore, the trajectories of the HVBs provide a promising way to constrain the IMBH orbital plane, if one exists. More importantly, we find that the ejected tight binaries are eccentric. For other galaxies hosting equal-mass SMBH binaries for which the HVB ejection rate is much higher relative to SMBH-IMBH binaries, these ejected compact binaries constitute promising extragalactic merger sources that could be detected by gravitational wave detectors. } 

\end{abstract}


\section{Introduction}
Hypervelocity stars (HVSs), ejected as a result of dynamical interactions with the supermassive black hole (SMBH) in the Galactic Centre (GC), are often traveling at speeds that exceed the Galactic escape velocity.  These remarkable objects were first predicted by \citet[][]{1988Natur.331..687H}. In this scenario, the tidal disruption of a tight stellar binary occurs as it passes close to a single massive black hole in the Galactic centre.  Alternative mechanisms for producing HVSs involve the scattering of a single star by a massive black hole binary  \citep[e.g.,][]{2003ApJ...599.1129Y,2006ApJ...651..392S},\yw{ dynamical evolution of a disk orbiting the SMBH \citep[e.g.,][]{2009ApJ...706..925B},  tidal interactions of star clusters with the SMBH \citep[e.g.,][]{2006ApJ...641..319P,2009ApJ...691L..63A}, supernova explosions \citep[e.g.,][]{1961BAN....15..265B,1970Natur.225..247C}, as well as the possibility that they may have originated in other galaxies, like the Large Magellanic Cloud \citep[e.g.,][]{2010ApJ...719L..23B}}. HVSs not only probe the gravitational potential of the Galaxy as they travel from the GC to the outer reaches of the Milky Way (MW), they also convey useful information about the environment(s) in which they form \citep[e.g.,][]{2006MNRAS.372..174B,2007MNRAS.379L..45S}. 

Since the discovery of the first HVS, a B-type main-sequence star traveling out of the Galaxy at twice the Galactic escape velocity \citep[e.g.,][]{2005ApJ...622L..33B}, many more HVSs have been found in the halo of the MW \citep[e.g.,][]{2006ApJ...647..303B, 2014ApJ...787...89B,2014ApJ...785L..23Z,2015ApJ...804...49B,2018MNRAS.479.2789B,2018MNRAS.476.4697M}. However, a few HVSs, such as HE0437-5439, a B-type main-sequence star whose travel time is longer than its main sequence lifetime, may not have been produced as a result of the standard Hills mechanism \citep[][]{2005ApJ...634L.181E}. \yw{Although \citet{2018arXiv180410197E} has shown that probably HE0437-5439 originated in the LMC, the previous hypothesis suggested by \citet{2010ApJ...719L..23B} (and originally discussed in \citealt{2009ApJ...698.1330P}) that this B-type HVS originated from a hypervelocity binary (HVB) ejected from the GC, that later evolved into a rejuvenated blue straggler, still remains debatable.}

{Another potential HVB has been identified by \citet{2016ApJ...821L..13N}. The wide orbit, of the order of an AU, however, suggests that the system might be associated with a marginally bound stellar stream rather than having been ejected from the MW centre. The argument is that such a wide binary would likely be disrupted in a close interaction rather than be ejected, but a thorough investigation of this possibility needs to be performed.} 
 
Several authors \citep[e.g.,][]{2007ApJ...666L..89L,2009MNRAS.392L..31S}
have investigated mechanisms to produce HVBs, in particular 
tight binary stars ejected by a massive black hole binary without being tidally torn apart. \yw{This could occur if there is an IMBH orbiting around Sgr A*, possibly delivered to the Galactic Centre by the infall of a globular cluster \citep[][]{2006ApJ...641..319P,2018arXiv180608385F, 2018ApJ...856...92F} or formed in situ in a disc surrounding Sgr A* \citep[][]{2012MNRAS.425..460M}}. Hypervelocity binaries cannot form in the standard Hills mechanism. However, HVBs may not be unique to SMBH binaries.  Close encounters between hierarchical triples/quadruples and a single SMBH\citep[e.g.,][]{2017MNRAS.467..451F,2018MNRAS.479.2615F,2018MNRAS.475.4986F} can also disrupt the hierarchical multiples and in so doing eject HVBs.

Assuming that 10\% of all interacting stars are triples, \cite{2018MNRAS.475.4986F} estimated the HVB rate from the tidal breakup of hierarchical multiples to be $<1$/Gyr, which is observationally negligible. For scattering of stellar binaries by an SMBH-IMBH binary, \citet[][]{2009MNRAS.392L..31S} showed that the rapid inspiral of a $5\times 10^4 M_\odot$ IMBH would generate $\sim 40$ HVBs before SMBH-IMBH coalescence. However, they did not investigate in detail the dependence of the ejection rate on individual scattering parameters; moreover, they did not follow the detailed dynamics of the stellar binary during the interaction or include finite sizes for the objects, thus neglecting the possibility of stellar mergers, which are likely for very compact binary systems. The interesting question that remains to be addressed is: what is the most probable region in the parameter space for the SMBH-IMBH scenario to eject HVBs, and what are the properties of these ejected HVBs?

To answer these questions, detailed scattering experiments between tight stellar binaries and an SMBH binary are needed \citep[][]{2018MNRAS.475.4595W}, and these require an extremely high-precision integrator. The close encounter of a stellar binary and an SMBH binary involves energy and angular momentum exchange among three orbits with a large energy difference: The inner stellar binary orbit has total energy $E_* \sim -Gm_*^2/2a_*$, the centre of mass orbit of the stellar binary has total energy $E_\cm\sim -Gm_*m_{\bullet 1}/2a_\cm$, and the SMBH-IMBH binary orbit has total energy $E_\bh\sim -Gm_{\bullet 1} m_{\bullet 2}/2a_\bh$, where $m_*$, $m_{\bullet 1}$, $m_{\bullet 2}$ indicate the mass of the star, the most massive SMBH, and the secondary IMBH, respectively. The parameters $a_*$, $a_\cm$, $a_\bh$ represent the semi-major axes of the stellar binary, the centre of mass orbit of the stellar binary and the SMBH-IMBH binary, respectively. For a typical stellar mass $m_*\sim m_\odot$ and assuming a mass for Sgr $A^*$ $m_{\bullet 1}\sim 4\times 10^{6}m_\odot$ along with an IMBH mass $m_{\bullet 2}\sim 4\times 10^{3}m_\odot$, the ratios between $E_*$, $E_\cm$ and $E_\bh$ are $6\times10^{-11}{a_\bh}/{a_*}$ and $2.5\times10^{-4}{a_\bh}/{a_\cm}$, respectively. This requires the total energy error of the whole system to be several orders of magnitude smaller than $6\times10^{-11}{a_\bh}/{a_*}$.  If this condition is not satisfied, any calculation of the energy and angular momentum transferred between the three orbits will be dominated by numerical errors. This numerical requirement becomes more rigid for smaller values of $a_\bh$, which is constrained by observational data.

In this paper, we explore the HVB ejection rate scenario described above using high-precision N-body integrations, with a focus on the observationally allowed parameter space for the orbital properties of a putative SMBH-IMBH binary in the MW centre. We investigate the energy and angular momentum exchanged during close encounters, and the properties of the resulting HVBs.

Our paper is organized as follows. In Section 2, we introduce the geometry of the four-body system, the initial set-up, and our numerical methods. In Section 3, we explore the relevant parameter space to find the highest-probability region to produce hypervelocity binaries. In Section 4, we study the interactions between the stellar binary and the IMBH, and the resulting properties of the hypervelocity binaries. Our findings are summarized in Section 5.

\section{INITIAL MODELS AND NUMERICAL METHODS}

\subsection{Initial models}
\begin{figure}
\centering
\includegraphics[width=0.8\columnwidth]{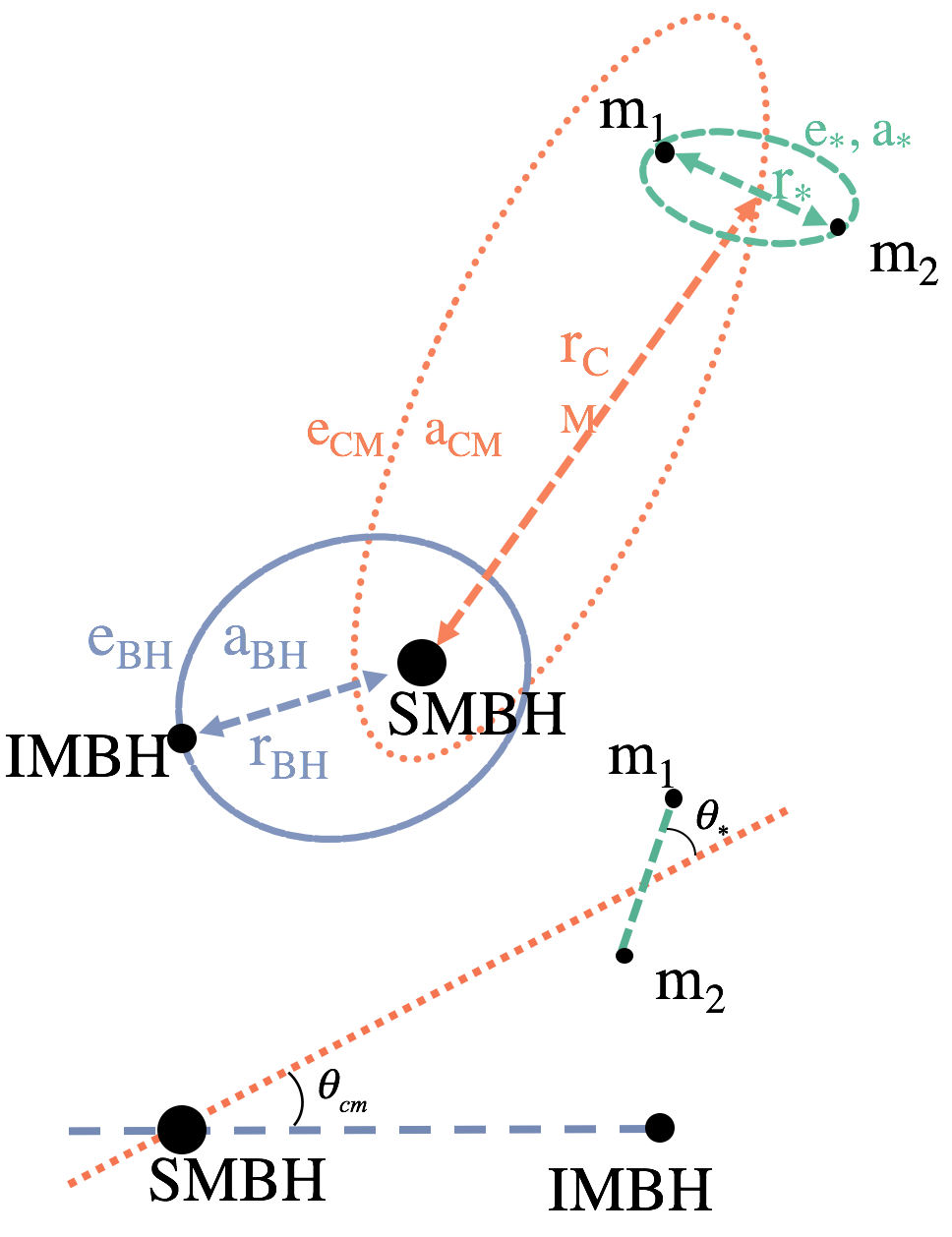}\\
\caption{Schematic illustration of the stellar binary orbiting around the super-massive black hole binary. }
 \label{fig:scheme}
\end{figure}

As shown in Figure (\ref{fig:scheme}), we label the masses of the stars by $m_{* 1}$ and $m_{* 2}$ (i.e., the stellar binary components), the mass of the primary SMBH by $m_{\bullet 1}$, and the mass of the secondary SMBH (here taken to be an IMBH) by
$m_{\bullet 2}$. For the orbital parameters, we use $a_{*,\cm,\bh}$ to denote the semi-major axes of the orbits, $e_{*,\cm,\bh}$ the eccentricities and $r_{*,\cm,\bh}$ the separations between the two components of each binary.  Here, the subscripts '$*$, $\cm$, $\bh$' denote, respectively, the (inner) orbit of the stellar binary, the orbit of the stellar binary about the primary SMBH, and the orbit of the secondary SMBH about the primary SMBH. 

\citet[][]{2003ApJ...599.1129Y} investigated the production of single hypervelocity stars via scatterings between single stars and massive black hole binaries. Based on the allowed region for the existence of an IMBH in the Galactic Centre, they derived constraints on the semi-major axis and the mass ratio of any potential SMBH-IMBH binaries. \yw{Those constraints were subsequently updated by \citet{2009ApJ...705..361G} using the long-term evolution of an S-star bound to Sgr A* in the Galactic centre.}  Here we adopt a mass ratio of 0.001 and a semi-major axis of $2\times 10^{-4}$~\pc, which are allowed by current observational limits. Thus, we choose the mass of the primary SMBH $m_{\bullet 1}$ to be $4\times 10^6 M_{\odot}$, and the mass of the secondary IMBH $m_{\bullet 2}$ to be $4\times 10^3 M_{\odot}$. The masses of the stars in the stellar binary are both set to $1M_{\odot}$.  {For such an SMBH-IMBH binary, the gravitational wave radiation timescale for a circular orbit is given by \cite{1964PhRv..136.1224P}
\begin{equation}
\tau_{GW} \sim \frac{5}{256}\frac{c^5a_\bh^4}{G^3m_{\bullet 1}m_{\bullet 2}(m_{\bullet 1}+m_{\bullet 2})}\sim 15~{\rm Myrs}\,.
\end{equation}
} 
This is sufficiently long relative to a Hubble time that we might still expect a timely merger event in the near future, if an IMBH exists.  

The initial conditions of our four-body problem are then completely defined by ten configuration parameters, which are shown in Table (\ref{tab:params}).

\begin{table*}
\begin{tabular} { | c | l | c | c | }
\hline
 Parameter & Description & Isotropic & Coplanar\\
 \hline
 $a_\bh$ &The semi-major axis of the black hole binary  & $2\times10^{-4} \pc$ & $2\times10^{-4} \pc$\\  
 $e_\bh$  & The eccentricity of the black hole binary & [0,1] & [0, 1]\\ 
 $l_\bh$ & The longitude of the second black hole's ascending node & $[0,2\pi]$ & $[0,2\pi]$\\ 
 $\theta_\cm$ & The inclination between the black hole binary orbit and the binary star orbit & $[0,\pi]$  & $0 \& \pi$ \\ 
 $a_{\cm}$ & The semi-major axis of the centre of mass of the stellar binary &$ [0.2, 3] a_{BH}$& $[0.2, 3]a_{BH}$\\ 
 $e_{\cm}$ &The eccentricity of the centre of mass of the stellar binary & $[0,1]$ & $[0,1]$\\ 
 $l_\cm$ &The longitude of the centre of mass of the stellar binary's ascending node & $[0,2\pi]$ & $[0,2\pi]$\\ 
 $\theta_*$ &The inclination of the binary star's inner orbit relative to its centre of mass orbit & $[0,\pi]$  & $0 \& \pi$\\ 
 $a_{*}$ &The semi-major axis of the stellar binary & $[0.023, 0.088] au$ & $[0.023, 0.088] au$\\ 
 $e_{*}$ &The eccentricity of the centre of mass of the stellar binary& 0 & 0\\ 
 $l_*$ &The longitude of the stellar binary's ascending node & $[0,2\pi]$ & $[0,2\pi]$\\ 
  \hline
\end{tabular}
\caption{Description of the initial configuration parameters. The square brackets indicate an uniform distribution between the enclosed values. }
\label{tab:params}
\end{table*}

For an isotropic stellar distribution, we sample $\cos\theta_\cm$ and $\cos\theta_*$ randomly in the range [-1,1], and $l_\bh$, $l_\cm$, $l_*$  randomly in the range [0,2$\pi$]. The parameter $a_{\cm}$ is sampled randomly in the range [0.2, 3]$a_{\bh}$, while $a_{*}$ is sampled randomly in log-space $[a_{\min},a_{\max}]$, where 
\begin{equation}
a_{\min} = 5 R_* = 5R_\odot (m_*/m_\odot)^{0.75},\;\;\; a_{\max}=\frac{Gm_*}{\sigma^2}
\end{equation}
with $R_*$ and $m_*$ representing the radius and the mass of the star, respectively, and $\sigma=100~$km~s$^{-1}$ being the velocity dispersion of the surrounding nuclear star cluster.  
Here, we use the the mass-radius relation $R_*/R_\odot = (m_*/m_\odot)^{0.75}$ \citep[e.g.,][]{2004sipp.book.....H}.

\subsection{Numerical Method}
We use $N$-body simulations to study the evolution of main-sequence binaries in the Galactic Centre (however, as discussed later, we emphasize that our results are just as applicable to extragalactic nuclear clusters). In each simulation, a stellar binary orbits the central SMBH with an outer IMBH/SMBH binary companion as shown in Figure (\ref{fig:scheme}). The equation of motion is determined by the Newtonian gravitational acceleration,
\begin{equation}
\ddot{\textbf{r}}_i = - \sum_{j\neq i} \frac{G m_j(\textbf{r}_i -  \textbf{r}_j)}{|\textbf{r}_i - \textbf{r}_j|^3}\,,
\end{equation}
where $i=1,2,3,4$. All simulations were carried out using our developing code \mbox{\href{https://github.com/YihanWangAstro/Template-SpaceX}{\tt SpaceHub}}. In this work, we use the {\tt ARCHAIN} algorithm \citep[][]{2008AJ....135.2398M}. This algorithm employs a regularized integrator to accurately trace the motion of tight binaries with arbitrarily large mass ratios, and a chain structure to reduce the round-off errors from close encounters. The Bulirsch-Stoer method \citep[e.g.,][]{1972BAAS....4T.422S} is embedded in this algorithm, to automatically choose the step size and the order of the integration needed to satisfy the given error tolerance using the lowest possible computing resources. We implement the {\tt ARCHAIN} algorithm with a template technique in {\tt C++} to eliminate the performance lost from the virtual function implemented in most object-oriented programming. We also adopt the data-oriented design needed to improve the cach\'e hit rate of the CPU. Basically, template programming allows the compiler to do more aggressive inline expanding, while data-oriented design helps the modern compiler to do SMID (Single Instruction, Multiple Data) optimization.  Both of these features significantly improve the performance of the code (more technical details will be provided in Wang et al. 2018, in prep.).

In our simulations, the local relative error of all physical quantities is set to be $10^{-16}$, while the total relative energy fluctuation is controlled at the level of $10^{-13}$.

\subsection{Simulated event detections}\label{sec:criteria}

Several physical phenomena play important roles in our simulations\citep[e.g.,][]{2017MNRAS.466.3376L, 2018MNRAS.475.4595W}.
We set the following criteria to identify and discriminate among the various event outcomes:

\begin{enumerate}{}{}
\item\textit{TDEs:}
Due to their low densities, main-sequence stars are easily tidally disrupted in the vicinity of a massive black hole. Tidal disruption occurs for stars that approach the black holes more closely than $r_{*t}$,
\begin{equation}\label{eq:TDE}
r_{*t}\sim \bigg(\frac{m_{\bh}}{m_*}\bigg)^{1/3}R_*
\end{equation}
where $m_{\bh}$ is the { mass of the massive black hole}, $m_*$ is the mass of the star and $R_*$ is the radius of the star. 

\item\textit{Binary disruptions:}
Stellar binaries will be broken apart by massive black holes if they enter the tidal breakup radius $r_{\mathrm{bt}}$,

\begin{equation}\label{eq:rbt}
r_{\mathrm{bt}}\sim \left(\frac{m_{\bh}}{m_{1}+m_{2}}\right)^{1/3}r_*
\end{equation}
where $r_{*}$ is the separation of the two components in the stellar binary. 

\item\textit{Mergers:}
Stars in the binary can collide with each other when the eccentricity of the stellar binary is very high. Instead of using the sticky star approximation in its usual form, which defines a collision as having occurred when the radii of the stars overlap, we assume that a collision occurs when the \textit{stellar cores} overlap directly. In our simulations, we set the core radius to be $0.2R_*$. This choice is somewhat arbitrary, but we emphasize that the statistics of HVB production do not change significantly if we use instead the stellar radius in the sticky-star approximation. Instead, this approximation is meant to avoid the extreme sensitivity of the sticky-star approximation to the orbital eccentricity in the contact binary limit. 

\item\textit{Double ejections:}
The stellar binaries can be ejected by the massive black holes through the slingshot effect. In our simulations, we identify binary ejections when the stellar binary is $50\,a_{\bh}$ away from the centre of mass of the four body system. This guarantees that the trajectory of the centre of mass of a given stellar binary is roughly Keplerian, such that a positive total energy for the binary centre of mass clearly defines an ejection event. We calculate the binary ejection energy as:
 \begin{equation}
 \begin{split}
 E_{eject} = &- \frac{G(m_{* 1}+m_{* 2})m_{\bullet 1}}{|\mathbf{r}_\cm-\mathbf{r}_{\bullet 1}|}
 - \frac{G(m_{* 1}+m_{* 2})m_{\bullet 2}}{|\mathbf{r}_\cm-\mathbf{r}_{\bullet 2}|}\\
 &+\frac{m_{* 1}+m_{* 2}}{2}\mathbf{v}_{\cm}^2\,,
 \end{split}
 \end{equation}
where $\mathbf{r}_\cm$ and $\mathbf{v}_\cm$ is the position and velocity of the stellar binary centre of mass. We also define the binding energy of the stellar binary,

\begin{equation}
E_{bind} = \frac{m_{* 1}}{2}(\mathbf{v}_{* 1}-\mathbf{v}_\cm)^2 + \frac{m_{* 2}}{2}(\mathbf{v}_{* 2}-\mathbf{v}_\cm)^2 - \frac{Gm_{* 1}m_{* 2}}{|\mathbf{r}_{* 1}-\mathbf{r}_{* 2}|}\,.
\end{equation}
In our simulations, positive $E_{eject}$ and negative $E_{bind}$ at $50\,a_{\bh}$ define a double ejection event.

\item\textit{Single ejections:}
The disruption of the stellar binary can create a single ejection event, whereby only one of the stars becomes an HVS.  For the single ejection event, we calculate the total energy of the ejected star,
 \begin{equation}
 E_{i}=\frac{1}{2}m_i\mathbf{v}_i^2 - \sum_{j\neq i} \frac{G m_i m_j}{|\mathbf{r}_i-\mathbf{r}_j|}\,. 
 \end{equation}
In our simulations, positive $E_{i}$ at $50\,a_{\bh}$ defines a single ejection event.
\end{enumerate}

There are some invalid initial conditions in the parameter space we explored. These are cases for which the initial eccentricity is very large, and hence collisions/TDEs can happen at the simulation time $t=0$.  These initial conditions satisfy the event criteria listed above, but should not be taken into account in calculating the final event rates. All of the initial conditions corresponding to such invalid events will be excluded in the event rate calculations. 

The criteria above are all considered as termination conditions in each individual simulation, \yw{together with setting the maximum integration duration to 5000 yrs. This duration time is long enough compared to the period of the SMBHB that we adopted in the simulation. As will be discussed in the next section, most HVBs are produced in non-secular regions of the parameter space. Therefore, long term secular effects are negligible in our simulations}. In this work, we focus more on events resulting in single and double ejections.

\section{HVB rate dependence on scattering parameters}\label{sec:allspace}
The possible outcomes of a stellar binary orbiting around an SMBH-IMBH binary are numerous. If the semi-major axis of the centre of mass of the stellar binary is small enough compared to the semi-major axis of the SMBH-IMBH binary, the four body system forms two hierarchical triples. The Kozai effect in the inner triple ($m_1$-$m_2$-SMBH) can induce merger events in the stellar binary, while the Kozai effect in the outer triple (($m_1$,$m_2$)-SMBH-IMBH) increases the probability of close encounters between the stellar binary and the primary SMBH, which enhances the TDE and ejection rates significantly. However, if the semi-major axis of the stellar binary increases, then the hierarchical multi-body system will gradually be broken apart. Chaotic effects come in to play, making the interaction outcome more unpredictable. While some papers have shown that chaotic effects could also play an important role in generating eccentric orbits in multi-body systems \citep[][]{2009ApJ...697L.149C}, the dependence on the initial conditions is not quite clear. Thus, here, we explore the parameter space covering both the hierarchical and chaotic regimes, in order to study the dependence of the event rates on the initial configuration parameters.  In this way, we identify the ideal environments for producing hypervelocity binaries.

We perform 150,000 integrations, sampling the relevant parameter space defined by the initial conditions specified in Table~\ref{tab:params}.  We hence measure the rate of HVSs and HVBs and study their properties.

\subsection{Stability}\label{sec:stability}
The parameter space in our model can be divided into secular and non-secular regions. In the secular region, the four body-system forms a hierarchy where Kozai-Lidov oscillations could play an essential role in the subsequent orbital evolution. In the non-secular region, the orbit of the stellar binary \yw{ gradually becomes unstable. In the mildly hierarchical region, fast mergers can be driven by quasi-secular evolution \citep{2014MNRAS.439.1079A,2018arXiv180802030G}. More importantly, in the strongly non-secular region, the high probability for close interactions between the stellar binary and the secondary IMBH creates an ideal environment to produce ejected stars.} 

For a triple system where the inner binary has component masses $m_{in,1}$ and $m_{in,2}$ and the perturber is an SMBH with mass $m_{out}$, \yw{we adopt the stability criterion derived by \citet{1996ASPC...90..433K}, which is more suitable for massive perturbers,}

\begin{equation}
Y=\frac{a_{out}(1-e_{out})}{a_{in}(1+e_{in})} > Y_{K,crit}
\end{equation}
where,
\begin{equation}\label{eq:crit1}
Y_{K,crit} = \frac{3.7}{Q_{out}}-\frac{2.2}{1+Q_{out}} + \frac{1.4}{Q_{in}}\frac{Q_{out}-1}{Q_{out}+1}
\end{equation}
with $Q_{in}=[\max(m_{in,1}/m_{in,2})]^{1/3}$, and $Q_{out}=[(m_{in,1}+m_{in,2})/m_{out}]^{1/3}$.

For a triple system in which the inner binary ($m_{in,1}$, $m_{in,2}$) contains an SMBH, \yw{we adopt the stability criterion derived by \citet{2001MNRAS.321..398M}, which is more suitable for massive inner binaries:}

\begin{equation}
Y=\frac{a_{out}(1-e_{out})}{a_{in}(1+e_{in})} > Y_{M,crit}\,,
\end{equation}
where
\begin{equation}\label{eq:crit2}
\begin{split}
Y_{M,crit} &= \frac{3.3}{1+e_{in}}\bigg[ \frac{2}{3} \bigg( 1 +\frac{m_{out}}{m_{in,1}+m_{in,2}} \bigg) \frac{1+e_{out}}{(1-e_{out})^{1/2}} \bigg]^{2/5}\\
 & \times  (1-0.3I/\pi)\,,
\end{split}
\end{equation}
with orbital inclination $I$.

For a stellar binary orbiting around an SMBHB, there are two triples in the system. The inner triple consists of $m_{* 1}$, $m_{* 2}$ and $m_{\bullet 1}$, while the outer triple consists of $(m_{* 1} + m_{* 2})$, $m_{\bullet 1}$ and $m_{\bullet 2}$. In the outer triple, the criterion for stability is
\begin{equation}
Y_{out} = \frac{a_\bh(1-e_\bh)}{a_\cm(1+e_\cm)} > Y_{M,crit}
\end{equation}
with $e_{in}=e_\cm$, $m_{out}=m_{\bullet 2}$, $m_{in,1}=m_{* 1}+m_{* 2}$, $m_{in,2}=m_{\bullet 1}$ and $e_{out}=e_\bh$ in Equation (\ref{eq:crit2}). For the inner triple, the criterion for stability is,
\begin{equation}
Y_{in} = \frac{a_\cm(1-e_\cm)}{a_*(1+e_*)} > Y_{K,crit}
\end{equation}
with 
\begin{equation}
Q_{in} = \max\bigg( \frac{m_{* 1}}{m_{* 2}}, \frac{m_{* 2}}{m_{* 1}}\bigg)^{1/3}
\end{equation}
and 
\begin{equation}
Q_{out} = \bigg( \frac{m_{* 1}+m_{* 2}}{m_{\bullet 1}}\bigg)^{1/3}
\end{equation}
in Equation (\ref{eq:crit1}).

Based on the above stability criteria, the parameter space can be divided into four parts.
\begin{enumerate}{}{}
\item\textbf{Unstable:} If $Y_{in}<Y_{K,crit}$ and $Y_{out}<Y_{M,crit}$, both the outer and inner triples are unstable. Strong non-secular effects dominate the system evolution. The stellar binary is easily disrupted and many close interactions with the IMBH occur.

\item\textbf{Inner stable:} If $Y_{in}>Y_{K,crit}$ and $Y_{out}<Y_{M,crit}$, the inner triple will remain stable. The inner Kozai-Lidov cycles from the SMBH dominate the system evolution and can easily drive the stellar binary to merger. Due to the hierarchical structure of the inner triple, there are few SMBH TDEs in this region of parameter space, and we do not expect many binary disruption events. Thus, most stellar binaries will remain bound. On the other hand, since the outer triple is unstable, the IMBH will break the centre of mass orbit of the stellar binary. Close encounters with the IMBH can efficiently produce IMBH TDEs, IMBH HVSs and IMBH HVBs. 

\item\textbf{Outer stable:} If $Y_{in}<Y_{K,crit}$ and $Y_{out}>Y_{M,crit}$, the outer triple forms a hierarchical system while the inner triple becomes unstable. The stellar binary in the inner triple is easily disrupted due to strong interactions with the SMBH. This binary disruption can produce hypervelocity stars by the standard Hills mechanism or leave two single stars orbiting around the SMBH. Due to the hierarchical outer triple, the Kozai-Lidov oscillations can drive the two single stars to get tidally disrupted by the SMBH. Therefore, in this region of parameter space, HVSs from close interactions with the SMBH and single TDEs with the SMBH are common.

\item\textbf{Double stable:} If $Y_{in}>Y_{K,crit}$ and $Y_{out}>Y_{M,crit}$, both the inner and outer triples form hierarchical systems. In this region of parameter space, double Kozai-Lidov oscillations can dramatically enhance the eccentricity of the centre of mass orbit of the stellar binary as well as the eccentricity of the inner stellar binary, leading to high rates of SMBH single TDEs, SMBH double TDEs and mergers. 

\end{enumerate}

\begin{figure}
\centering
\includegraphics[width=\columnwidth]{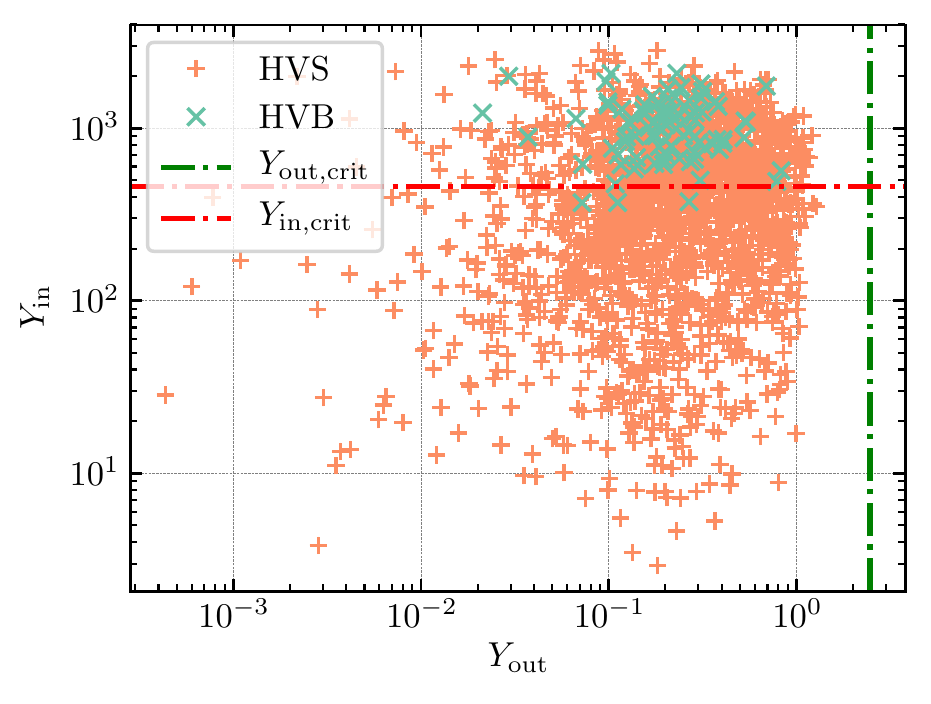}\\
\caption{HVSs and HVBs in the stability map $Y_{in}$ and $Y_{out}$. The red dashed-dotted line is the critical $Y_{in}$ to form a stable inner hierarchical triple. \yw{The $Y_{out,crit}$ of each single HVB/HVS event is calculated using Equation~(\ref{eq:crit2}) (but note that this value for  $Y_{out}$ is not a constant). The green line in the plot shows the average value of all $Y_{out,crit}$.} All HVSs and HVBs are produced in the outer unstable region where $Y_{out}<Y_{M,crit}$, with HVBs occurring almost entirely in the outer unstable-inner stable region where $Y_{out}<Y_{M,crit}$ and $Y_{in}>Y_{K,crit}$.}
 \label{fig:division}
\end{figure}

To produce an HVB, the two stars must remain bound in a binary, which requires the inner triple to be stable. On the other hand, the ejection needs a close encounter with the IMBH. Thus, the most promising region of parameter space to produce HVBs is the inner stable region where $Y_{in}>Y_{K,crit}$ and $Y_{out}<Y_{M,crit}$. Figure (\ref{fig:division}) shows the HVSs and HVBs produced in the $Y_{out}-Y_{in}$ region of parameter space. All the ejections (HVSs and HVBs) are produced in the region of parameter space for which a strong interaction occurs between the IMBH and the stellar binary ($Y_{out}<Y_{M,crit}$). More importantly, HVBs are effectively produced in the inner stable region where $Y_{in}>Y_{K,crit}$ and $Y_{out}<Y_{M,crit}$. This implies that the presence of an HVB could signal the existence of an SMBH-IMBH/SMBH binary.

In the following subsections, to further investigate the properties of HVBs, we will focus on the inner stable region of the parameter space.

\subsection{Orbital inclination}

\begin{figure}
\centering
\includegraphics[width=\columnwidth]{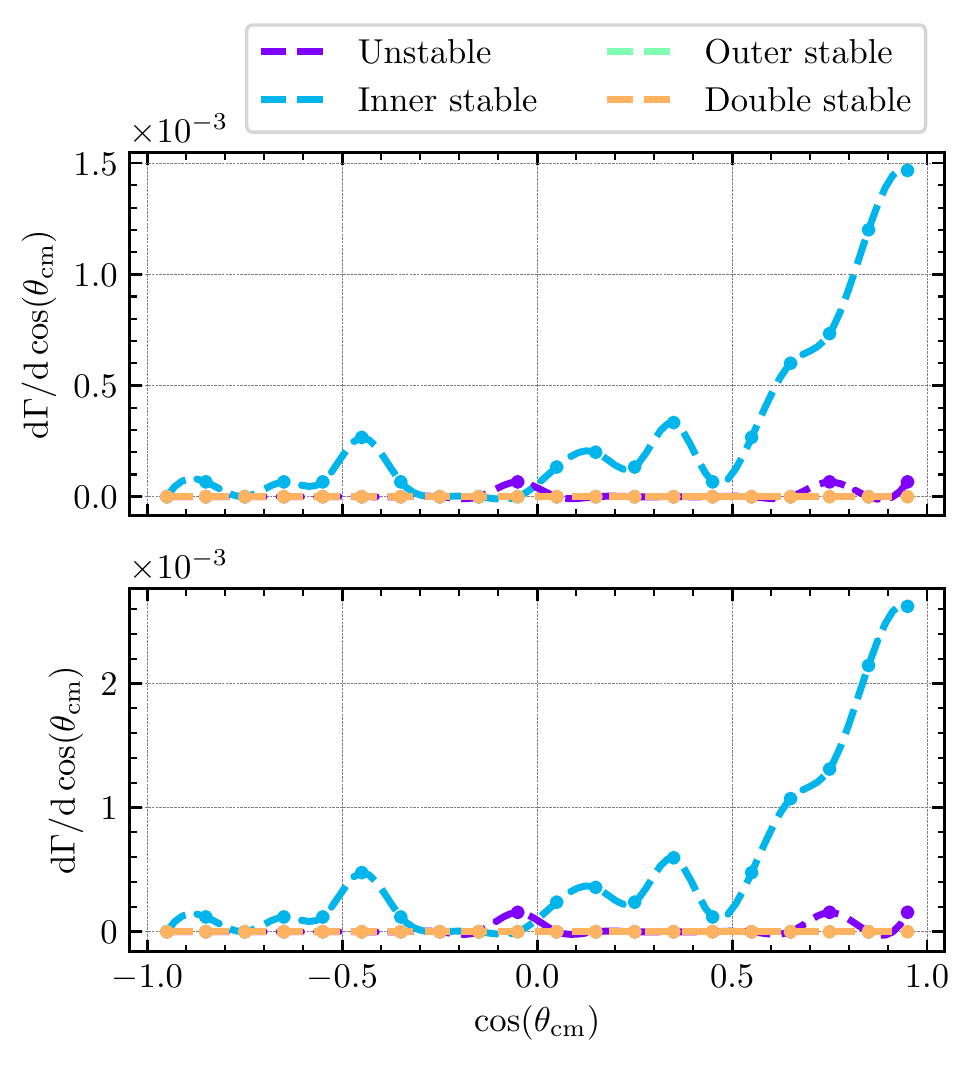}
\caption{HVB relative rate \yw{density ($\Gamma$)} as a function of $\cos\theta_\cm$ for four different channels in the stability map. The blue dotted-dashed line represents the inner stable channel, the yellow line indicates the double stable channel, the cyan line the unstable channel and the green line the outer stable channel. The various channels are described in Section (\ref{sec:stability}). Note that lines superimposed at 0.0 are not clearly visible. The upper panel is normalized by the total simulation sample volume while the bottom panel is normalized by the sample volume of different channels. Almost all HVBs come from the inner stable channel and from low inclination angles $\theta_\cm$ and prograde orbits.}
\label{fig:thetachoose}
\end{figure}
Based on the stability analysis of the last subsection, we have now formed a general idea of the relative event occurrence rates corresponding to different regions of the total parameter space. An interesting question concerns how the HVB relative event occurrence rate ratios depend on the orbital inclinations $\theta_\cm$ and $\theta_*$, especially in the inner stable region where $Y_{in}>Y_{K,crit}$ and $Y_{out}<Y_{M,crit}$. In this subsection, we study the statistics of HVBs formed in all four regions of parameter space, and show their dependencies on $\theta_\cm$ and $\theta_*$.

Figure (\ref{fig:thetachoose}) shows the dependence of the HVB rate on the initial inclination between the black hole binary orbit and the binary star orbit. The upper panel shows the relative event rates normalized by the total sample volume from all regions of parameter space, while the bottom panel shows the relative rates normalized by the sample volume of each region. As discussed in the stability section, almost all HVBs come from the inner stable region, which is indicated by the blue dash-dotted line. As seen in both panels, HVB production has a strong dependence on the outer triple inclination $\theta_\cm$.  HVBs are mostly produced when on prograde orbits with $\cos\theta_\cm > 0$ and are rarely associated with retrograde orbits with $\cos\theta_\cm < 0$. Since HVBs are created by close encounters with the IMBH, this phenomenon can be understood using cross-section theory: the cross-section for encounters with an IMBH on a prograde orbit is larger than for retrograde orbits. 

\yw{Let $R_{e}$ be the effective ejection radius and $v_e$ the corresponding velocity at this radius, and let the stellar binaries come in with centre of mass velocity $v_0$ with respect to the IMBH, and with impact parameter $b$. Due to energy and angular momentum conservation (and ignoring the small energy contribution from the gravitational potential energy at $b$),
\begin{equation}
\frac{1}{2}m_b v_0^2 = \frac{1}{2}m_b v_e^2 - {Gm_{\bullet2 }m_b}/{R_e},~
b v_0 = R_e v_e
\end{equation}
the effective ejected cross section is
\begin{equation}
    \sigma = \pi b^2 =\pi R_e^2\left(1+\frac{2Gm_{\bullet2 }}{R_e}\frac{1}{v_0^2}\right)\,.
\end{equation}
The relative velocity of the centre of mass of the stellar binaries $v_0$ is much larger in retrograde orbits, thus
}the HVB rate reaches its maximum value at $\theta_\cm = 0$, which corresponds to the prograde coplanar case.

\begin{figure}
\centering
\includegraphics[width=\columnwidth]{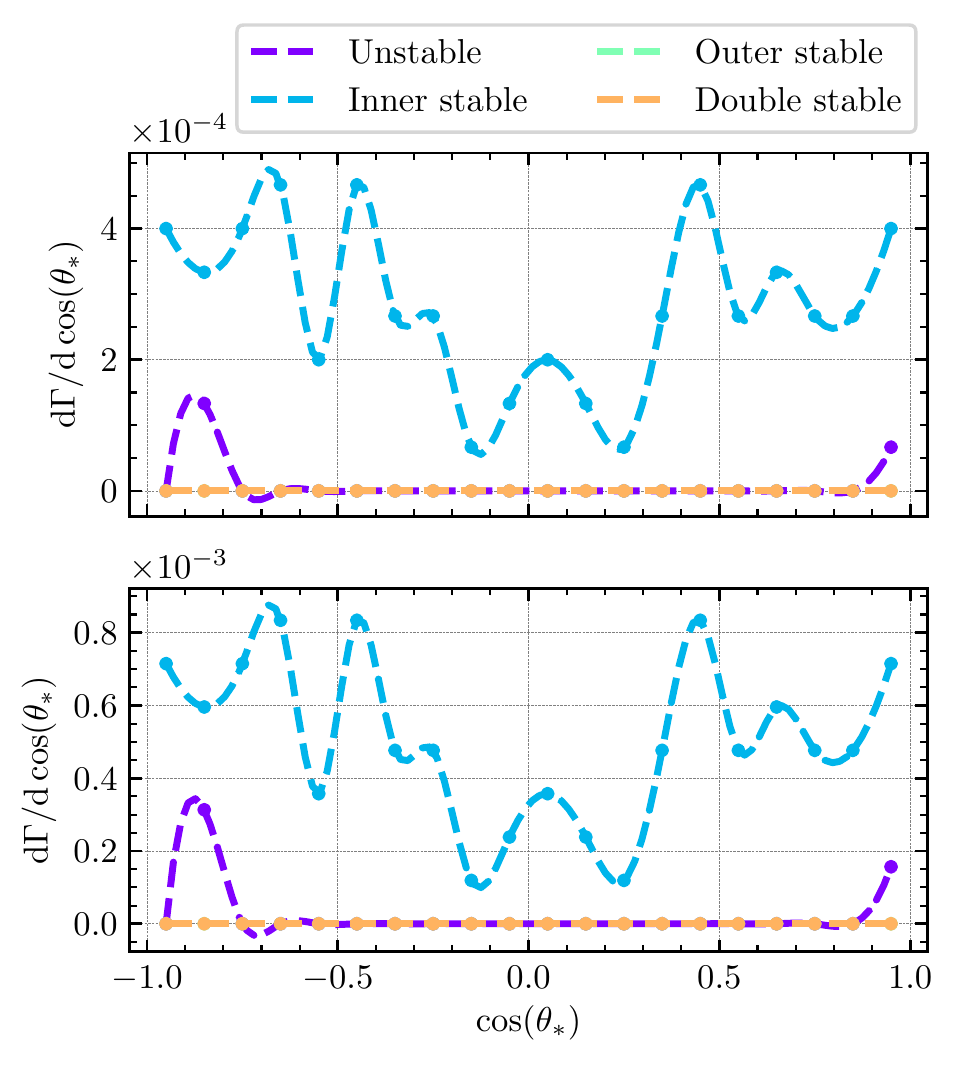}
\caption{HVB relative event occurrence rate \yw{density $(\Gamma$)} as a function of $\cos\theta_*$ for all four regions in the stability map. The upper panel is normalized by the total simulation sample volume while the bottom panel is normalized by the sample volume for each region. HVBs tend to be associated with low $\theta_*$ orbits where no inner Kozai-Lidov oscillations operate and hence no stellar binary mergers occur.}
\label{fig:thetaschoose}
\end{figure}

Figure (\ref{fig:thetaschoose}) shows the dependence of the HVB rate on the initial inclination of the binary star's inner orbit relative to its centre of mass orbit $\theta_*$. Similar to Figure (\ref{fig:thetachoose}), the upper and bottom panels show the relative rate normalized by the total sample volume, and by the sample volume for each region, respectively. HVB production is symmetric for prograde and retrograde orbits (i.e. there are no statistical differences between these two cases), and are likely to be produced at low inclination angles $\theta_*$. Since most HVBs are produced in the inner stable region of parameter space where the inner Kozai-Lidov oscillations play an essential role, high inclination $\theta_*$ orbits are easily driven to mergers by the inner Kozai-Lidov oscillations. {Therefore, the orbits at low $\theta_*$ have a larger probability of surviving and producing HVBs by close interactions with the IMBH }

\subsection{Semi-major axis of the stellar binary}
\begin{figure}\label{fig:astar}
\centering
\includegraphics[width=\columnwidth]{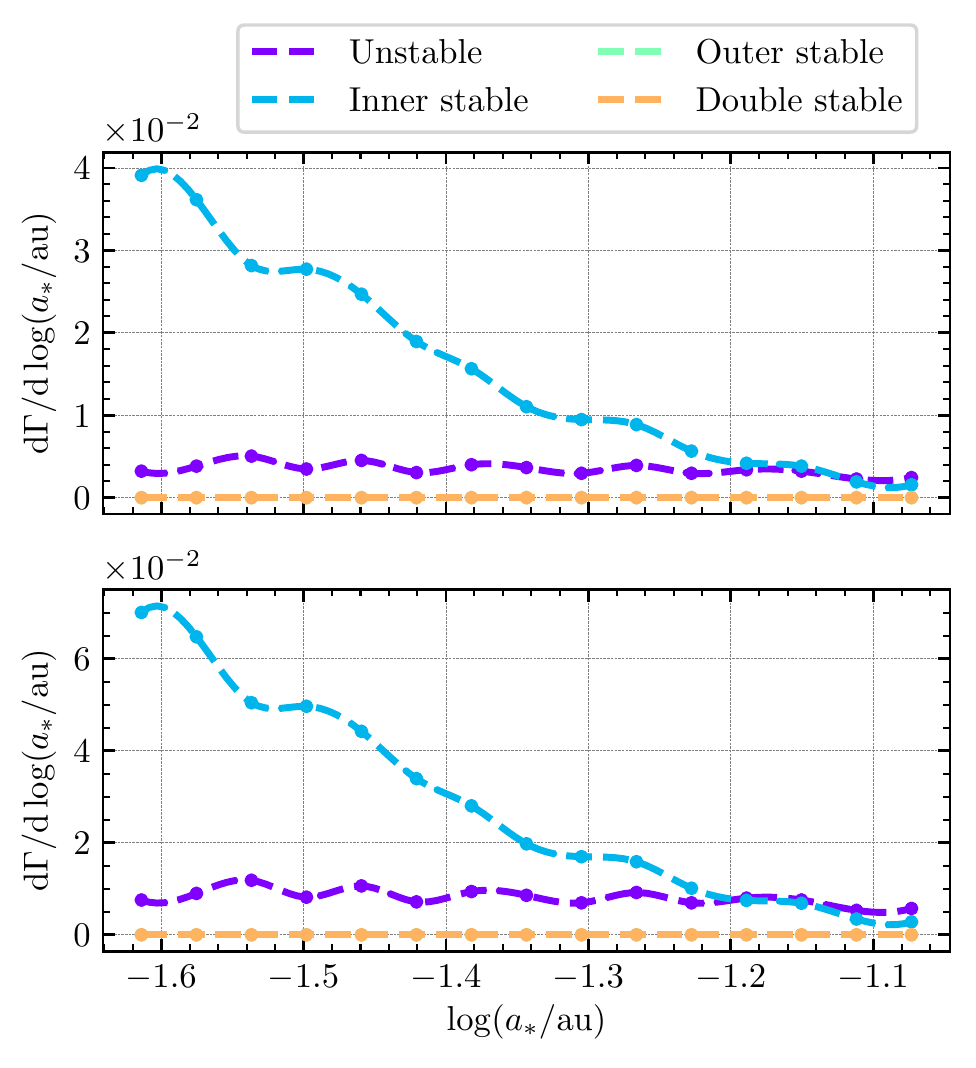}
\caption{HVB relative rate \yw{($\Gamma$) density} as a function of $a_*$ for all four regions in the stability map. The upper panel is normalized by the total simulation sample volume while the bottom panel is normalized by the sample volume of each region. More compact stellar binaries get ejected more easily.}
\label{fig:astarchoose}
\end{figure}

Based on the dependencies of HVB production on the orbital inclination and close interactions with the IMBH, we ran 300,000 integrations to further sample the parameter space defined in Table (\ref{tab:params}), assuming coplanar configurations. We sampled the semi-major axis of the MS binary $a_*$ evenly in log-space. 

Figure (\ref{fig:astarchoose}) indicates that compact binaries are easier to eject as HVBs in the coplanar case. Here we explore the coplanar case in the inner triple channel for HVB production, and there are no Kozai-Lidov oscillations in such a configuration. Most HVB events in this coplanar case are due to close interactions with the IMBH. Generally, compact binaries have a higher probability of surviving the close encounters with the IMBH, for a given semi-major axis and eccentricity of the binary centre of mass orbit. This is because compact binaries have a smaller disruption radius, as described in Equation (\ref{eq:rbt}). Therefore, we expect more HVBs from the most compact binaries, such as those consisting of compact objects. For more compact binaries such as neutron star and stellar BH binaries in which post-Newtonian corrections in the stellar binary could become important, a more detailed investigation will be performed in future work.

\section{Hypervelocity binaries produced in coplanar configurations}
Based on our analysis in Section (\ref{sec:allspace}), we find that the ideal environment to produce hypervelocity binaries corresponds to purely prograde coplanar configurations. In this section, we focus on HVBs produced in stellar disks, which generally satisfy this criterion to first-order. 

\subsection{Velocity kicks}
\begin{figure}
\centering
\includegraphics[width=\columnwidth]{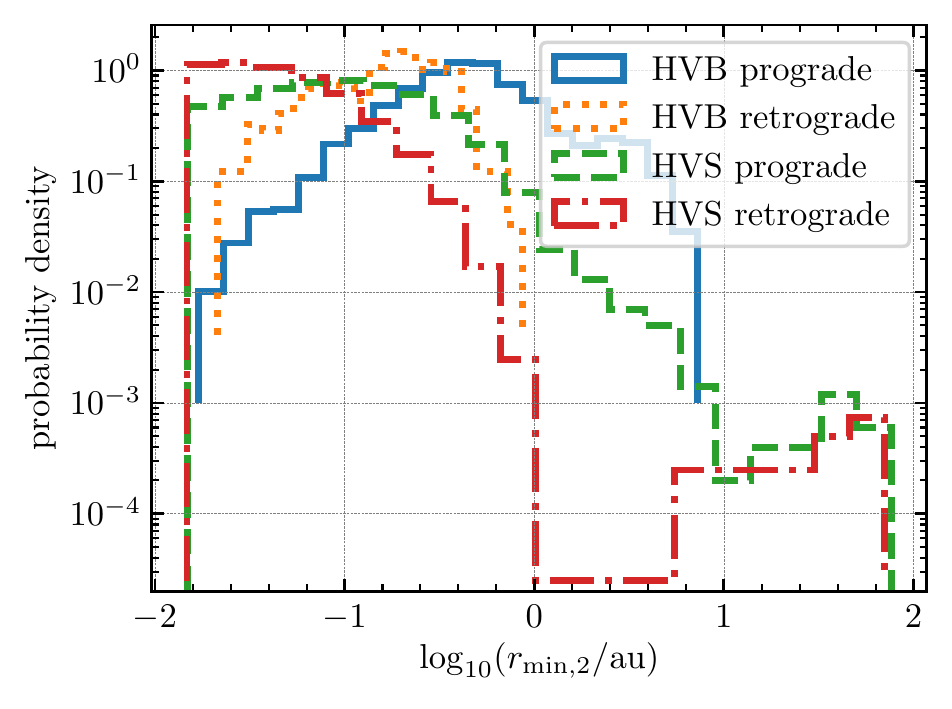}\\
\caption{\yw{The closest approach of the stellar binaries to the IMBH in both HVS and HVB events. Retrograde orbits can get closer to the IMBH in both HVS and HVB events due to the larger relative velocities between the stellar binaries and the IMBH.} }
\label{fig:closest}
\end{figure}

Close encounters between stellar binaries and the IMBH produce HVBs by the slingshot mechanism. Encounters between compact binaries and the IMBH are such that the resulting binary centre of mass velocity ${v}_\cm$ is of the same order as the IMBH orbital velocity, $v_\cm\sim v_{\bullet 2}=(Gm_{\bullet 1}/r_\bh)^{1/2}$. If the stellar binary has a close approach with the IMBH $r_{\min,2}$, it will receive a local velocity kick $\Delta v_\cm\sim(Gm_{\bullet 2}/r_{\min,2})^{1/2}$ due to the force imparted from the IMBH $\sim Gm_{\bullet 2}/r^2_{\min,2}$ with a corresponding encounter timescale of $\sim(r^3_{\min,2}/Gm_{\bullet 2})^{1/2}$\citep[e.g.,][]{1996NewA....1...35Q}. This leads to,
\begin{equation}\label{eq:v-q}
\frac{\Delta v_\cm}{v_{\bullet 2}}\sim \bigg( \frac{r_\bh m_{\bullet 2}}{r_{\min,2}m_{\bullet 1}}\bigg)^{1/2}=(\frac{qr_\bh}{r_{\min,2}})^{1/2}\,,
\end{equation}
where $q=m_{\bullet 2}/m_{\bullet 1}$. Figure~(\ref{fig:closest}) shows the closest approach of the stellar binaries to the IMBH of the HVSs and HVBs, both in prograde and in retrograde orbits. It can be seen that the HVSs can get closer to the IMBH than the HVBs, because the HVBs will get tidal disrupted if Equation (\ref{eq:rbt}) is satisfied. Additionally, we also find that the retrograde orbits get closer to the IMBH. Stars in retrograde orbits have larger relative velocities with respect to the IMBH during the close encounter; thus, they can reach deeper regions of the potential of the IMBH. 

\begin{figure}
\centering
\includegraphics[width=0.95\columnwidth]{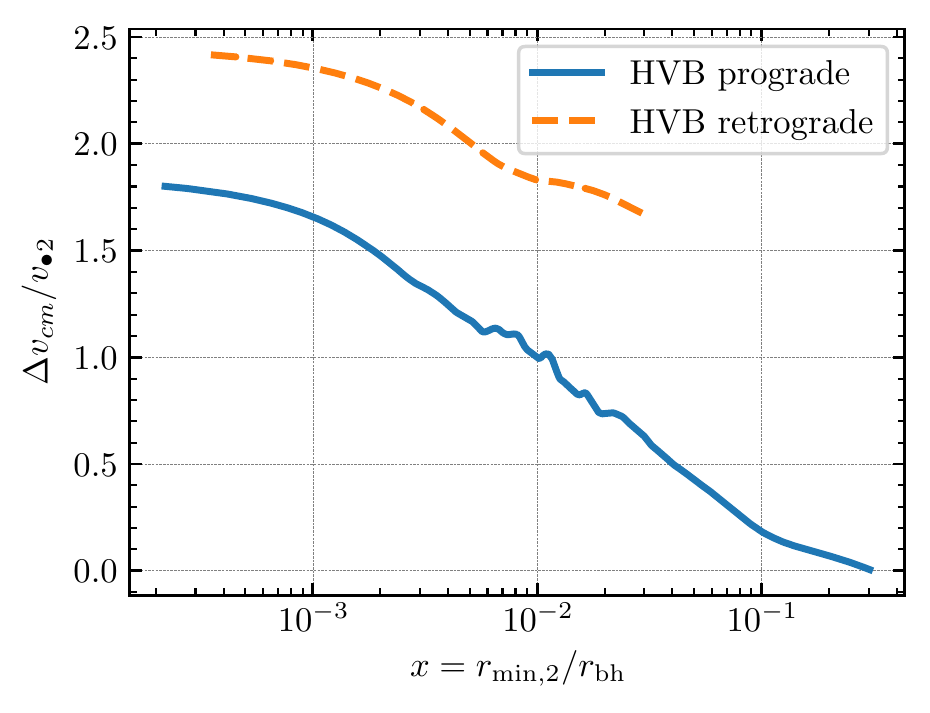}
\caption{Local mean kick velocity as a function of the closest approach distance to the secondary IMBH. \yw{The maximum mean velocity kick that the stellar binary can acquire from interacting with the IMBH is roughly twice the orbital velocity of the IMBH.}}
\label{fig:compare}
\end{figure}

Figure~(\ref{fig:compare}) shows the relationship between $\Delta v_\cm/v_{\bullet 2}$ and $x=r_{\min,2}/r_\bh$. As expected from Equation~(\ref{eq:v-q}), $\Delta v_\cm$ and $r_{\min,2}$ show a linear relationship in log-log space, as also found by \citet{2009MNRAS.392L..31S}. Given the fact that, considering the stellar binary as a single particle, Equation~(\ref{eq:v-q}) does not account for the energy and angular momentum transfer between the internal stellar binary and its centre of mass orbit,  it is then not surprising that the average velocity kicks of prograde and retrograde orbits are different, if the energy transfer between the internal binary and its centre of mass orbit are different. Here, we see a clear gap between the prograde and retrograde orbits. In fact, the retrograde close encounters can transfer more energy from the internal binary to its centre of mass orbit relative to prograde close encounters. Therefore, for the same closest approach distance, the retrograde encounters get a larger velocity kick from the IMBH. This finding is clearly illustrated in a later subsection which discusses the energy exchange during the close encounter, as indicated by the upper right panel of Figure~(\ref{fig:Ex}).

\subsection{Angular momentum exchange}

\begin{figure*}
\centering
\includegraphics[width=\columnwidth]{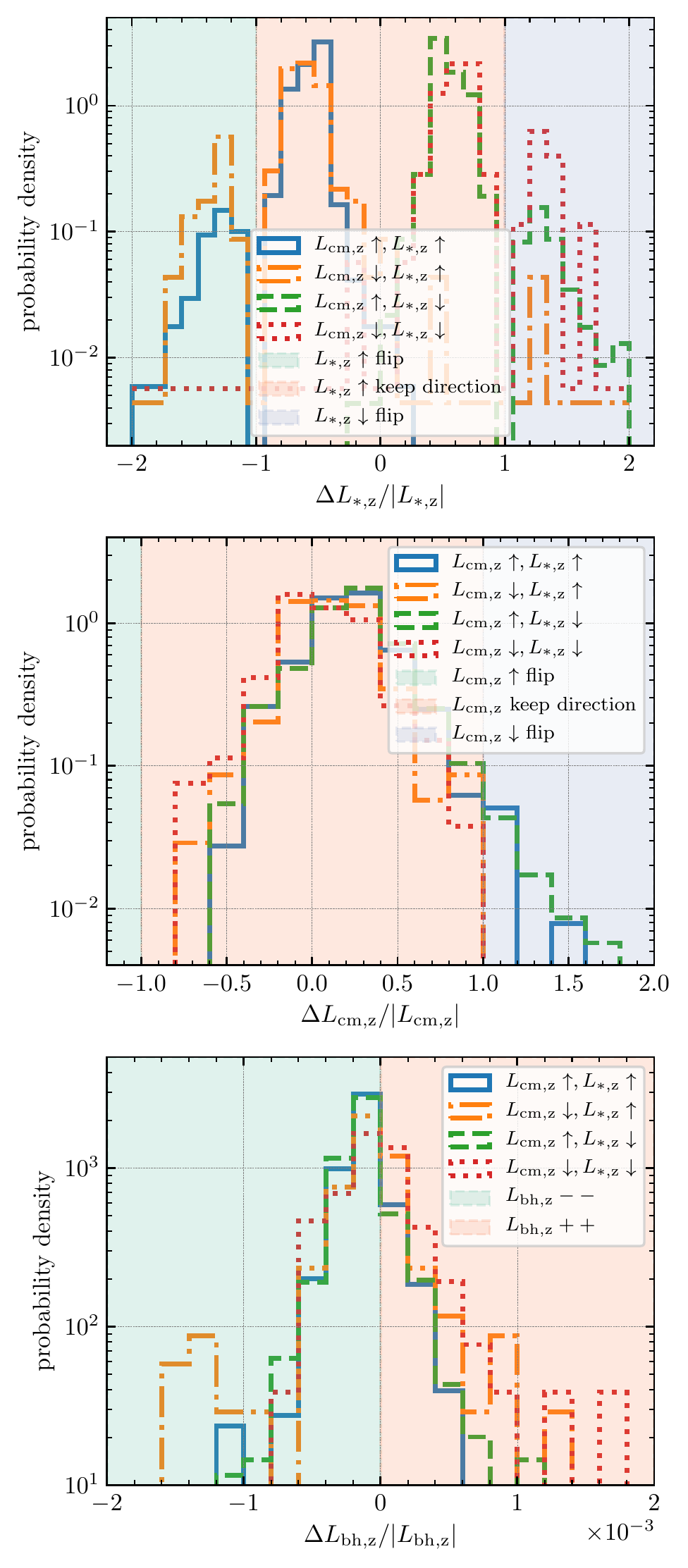}
\includegraphics[width=\columnwidth]{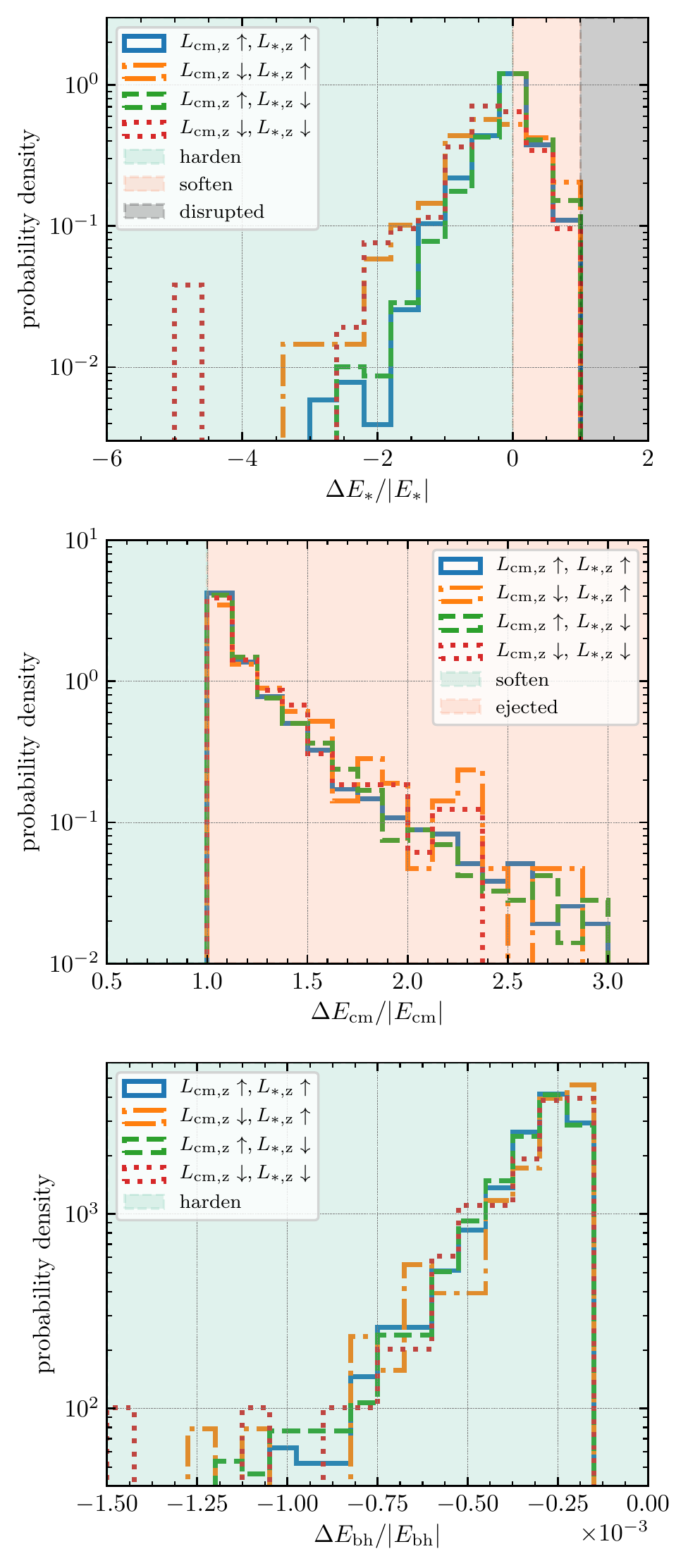}
\caption{The energy and angular momentum exchanged between the stellar binary, the centre of mass orbit of the stellar binary and the SMBH-IMBH binary orbit during close encounters. The upper left panel shows the change in angular momentum of the inner stellar binary post-ejection.  Orbits with $\mathbf{L}_{*,z}>0$ tend to lose angular momentum, while orbits with $\mathbf{L}_{*,z}<0$ tend to gain angular momentum. The upper right panel shows the change in energy of the stellar binary, which indicates that stellar binaries tend to be more compact after ejection. The bottom left panel displays the change in angular momentum of the centre of mass orbit of the stellar binaries. Prograde and retrograde orbits show different profiles due to the different relative velocity between the IMBH and the stellar binary. The bottom right panel shows the change in energy of the centre of mass orbit of the stellar binary. Since all HVBs escape from the SMBHB on hyperbolic orbits, all stellar binaries gain enough energy from the IMBH to make their orbits hyperbolic.}
\label{fig:Ex}
\end{figure*}

\begin{figure*}
\centering
\includegraphics[width=\columnwidth]{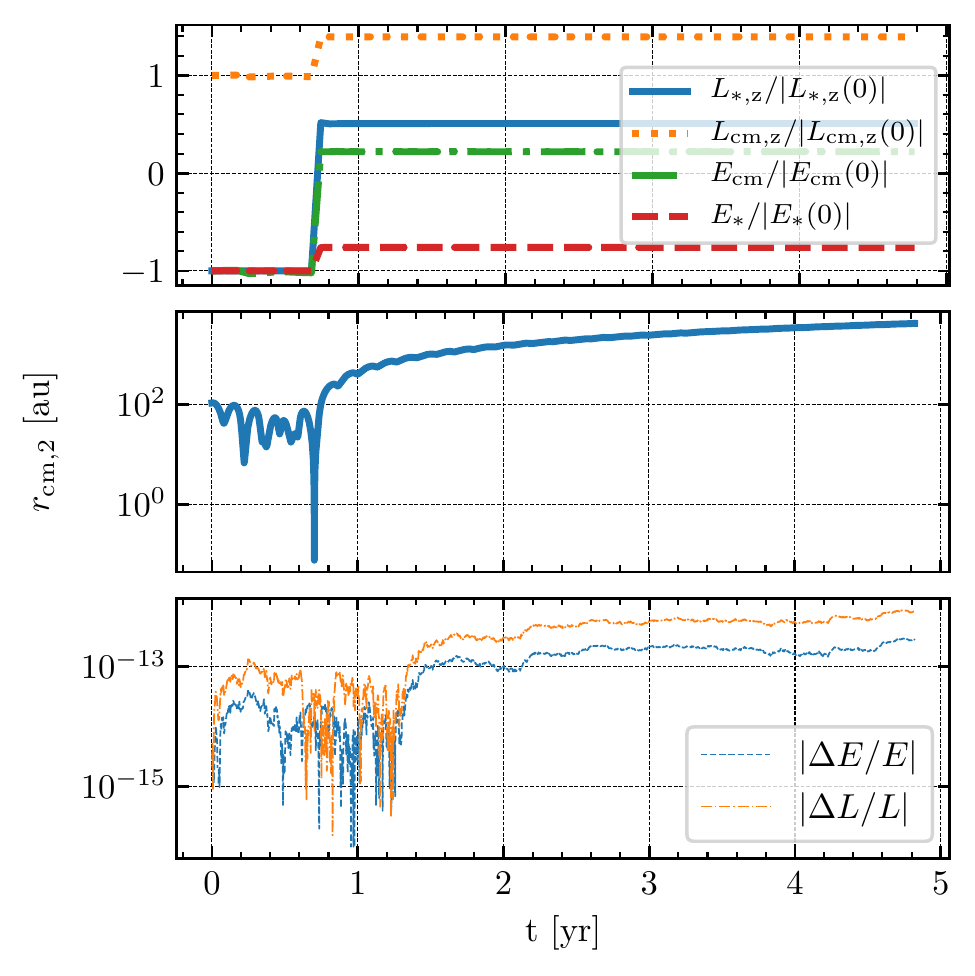}
\includegraphics[width=\columnwidth]{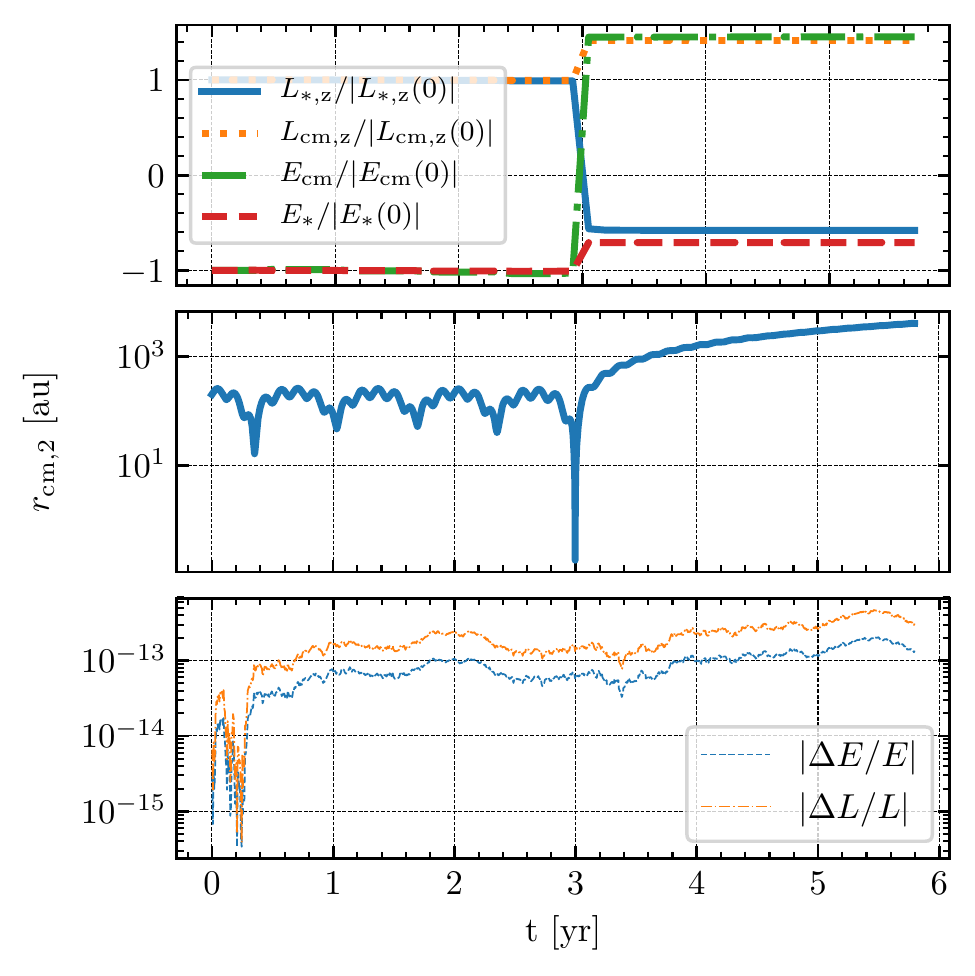}\\
\includegraphics[width=\columnwidth]{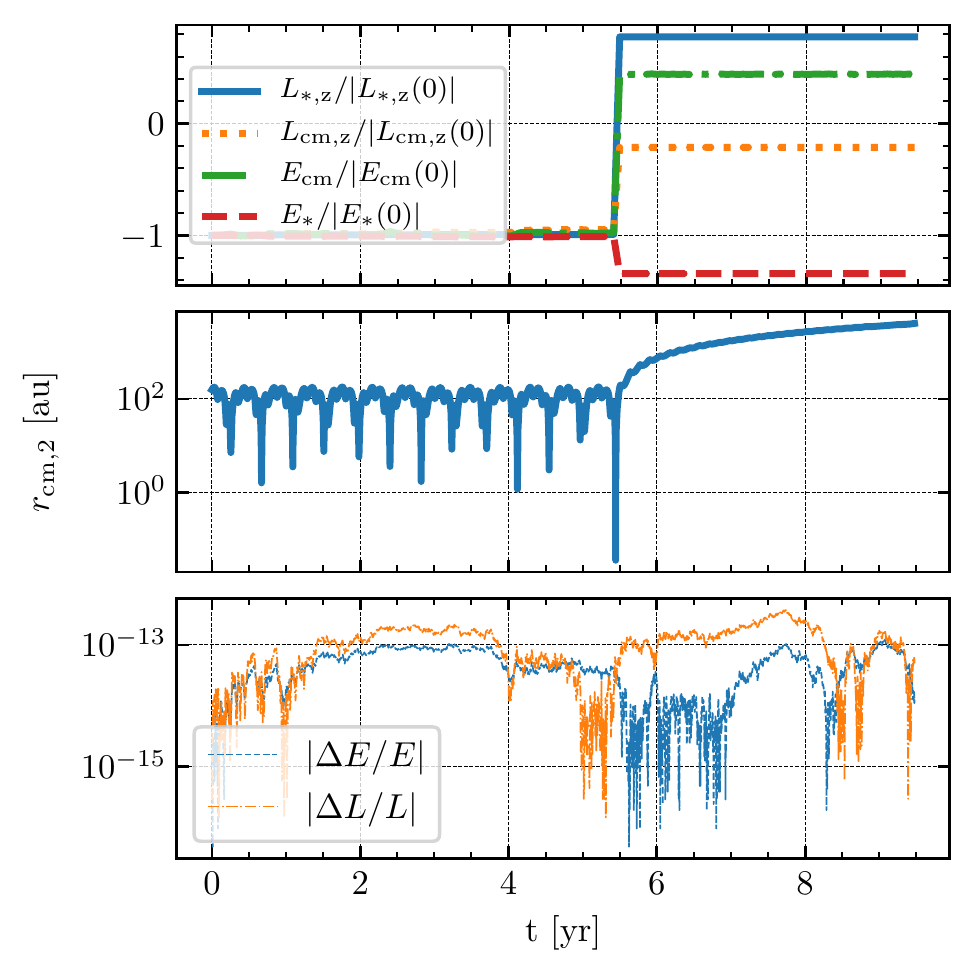}
\includegraphics[width=\columnwidth]{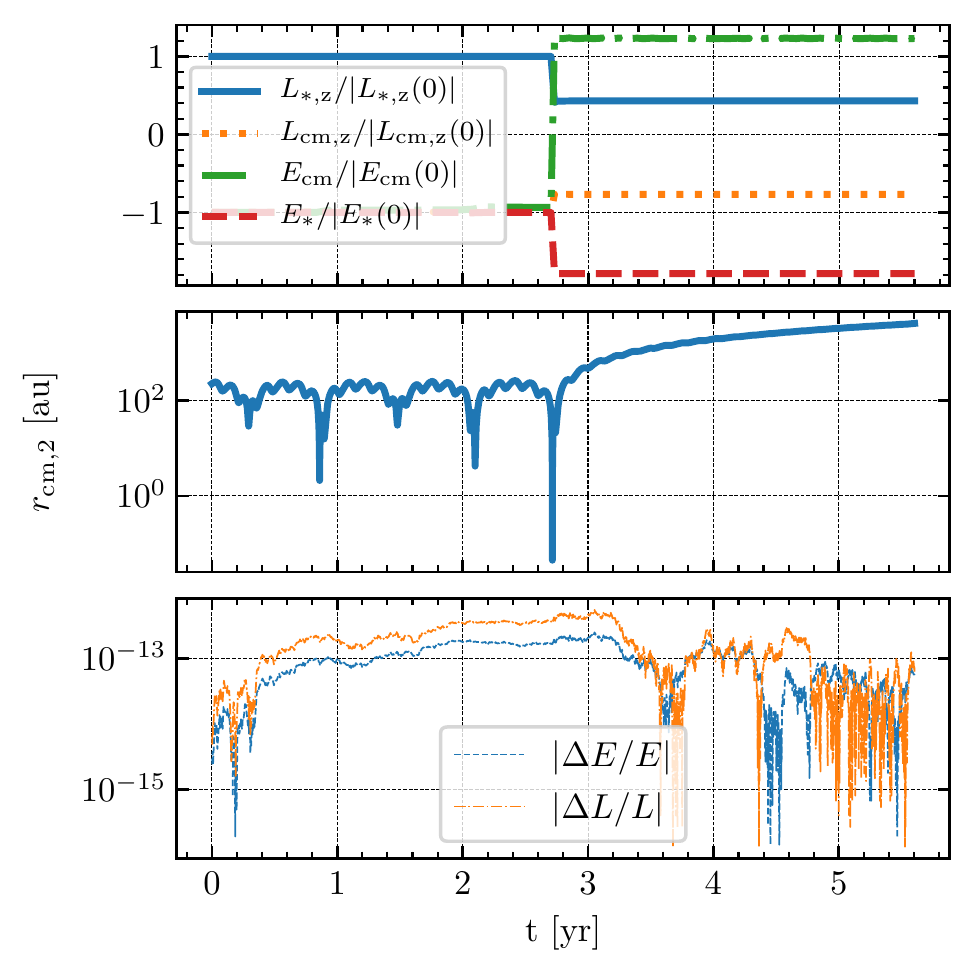}\\
\caption{Four examples of \yw{the HVB ejection process, for} different configurations. The upper left panel shows an example of the inner retrograde-outer prograde case. The upper right panel displays an example of the inner prograde-outer prograde case. The bottom left shows an inner retrograde-outer retrograde case. Last, in the bottom right, we show an example of the inner prograde-outer retrograde case.}
\label{fig:progInnerFlip}
\end{figure*}

The left two panels in Figure~(\ref{fig:Ex}) show the z-component of the angular momentum exchange before and after the close encounter. $L_*$ represents the angular momentum of the inner stellar binary and is equal to,
\begin{equation}
\begin{split}
\mathbf{L}_* &&= m_{* 1}(\mathbf{r}_{* 1}-\mathbf{r}_\cm)\times(\mathbf{v}_{* 1}-\mathbf{v}_\cm) \\
&&+ m_{2*,}(\mathbf{r}_{* 2}-\mathbf{r}_\cm)\times(\mathbf{v}_{* 2}-\mathbf{v}_\cm)\,,
\end{split}
\end{equation}
where $\mathbf{r}_{* 1},\mathbf{r}_{* 2},\mathbf{v}_{* 1},\mathbf{v}_{* 2}$ represent the position and velocity of $m_{* 1}$ and $m_{* 2}$, respectively. $\mathbf{L}_\cm$ is the angular momentum of the centre of mass orbit of the stellar binary, which is equal to:
\begin{equation}
\mathbf{L}_\cm = (m_{* 1}+m_{* 2})\mathbf{r}_\cm\times \mathbf{v}_\cm
\end{equation}
where $\mathbf{r}_{\cm}$ and $\mathbf{v}_{\cm}$ represent, respectively, the position and velocity of the centre of mass of the stellar binary. {
$\mathbf{L}_\bh$ is the angular momentum of the SMBH-IMBH binary, which is equal to:
\begin{equation}
\mathbf{L}_\bh = m_{\bullet 1}\mathbf{r}_{\bullet 1}\times \mathbf{v}_{\bullet 1} + m_{\bullet 2}\mathbf{r}_{\bullet 2}\times \mathbf{v}_{\bullet 2}\,.
\end{equation}
}
The change in angular momentum $\Delta \mathbf{L}_z$ is calculated as the difference between the initial and final (i.e., post-ejection, evaluated far away from the SMBH-IMBH binary, at $\sim 50~a_\bh$) angular momenta of the binary. The upper left panel of Figure~(\ref{fig:Ex}) indicates that almost all of the orbits with $\mathbf{L}_{*,z}>0$ lose angular momentum during the close encounter with the IMBH, while having a symmetric distribution, whereas almost all of the orbits with $\mathbf{L}_{*,z}<0$ gain angular momentum during the close encounter with the IMBH. This means that the internal stellar binary tends to change its angular momentum direction. For those HVBs for which $\Delta \mathbf{L}_{*,z}/|\mathbf{L}_{*,z}| > 1$ with $\mathbf{L}_{*,z} < 0$ and $\Delta \mathbf{L}_{*,z}/|\mathbf{L}_{*,z}| < -1$ with $\mathbf{L}_{*,z} > 0$, the final direction of $\mathbf{L}_{*,z}$ gets flipped. Moreover, the retrograde orbits with $\mathbf{L}_{\cm,z} < 0$ have more opportunities to get flipped, due to the stronger interactions in such encounter configurations. 

The middle left panel of Figure~(\ref{fig:Ex}) shows the angular momentum change of the stellar binary centre of mass orbit. It is clear that no orbit gets flipped after the ejection, which means the close encounters cannot change the prograde or retrograde nature of the centre of mass orbit of the stellar binary. More importantly, we find that the retrograde orbits with $\mathbf{L}_{\cm,z}<0$ are more likely to lose angular momentum than the prograde orbits with $\mathbf{L}_{\cm,z}>0$. The difference between $\mathbf{L}_{*,z}>0$ and $\mathbf{L}_{*,z}<0$ with the same $\mathbf{L}_{\cm,z}$ direction is very small here. If we consider a possible difference here, due to the difference in angular momentum $\Delta \mathbf{L}_{*,z}$ with $\mathbf{L}_{*,z}$ being either positive or negative as shown in the top left panel, we might expect identical distributions but in opposite directions, and we would see a gap here at the order of magnitude level,
\begin{equation}
\frac{|\Delta \mathbf{L}_{*,z}|}{|\Delta \mathbf{L}_{\cm,z}|}\sim \frac{|\mathbf{L}_{*,z}|}{|\mathbf{L}_{\cm,z}|}\sim\bigg(\frac{a_*\mu_*}{a_\cm\mu_\cm}\bigg)^{1/2}\,,
\end{equation}
where the reduced mass of the stellar binary is $\mu_*$ and the reduced mass of the centre of mass orbit of the stellar binary is $\mu_\cm$. This is because both $\mathbf{L}_{cm,z}$ and $\mathbf{L}_{*,z}$ change at the same order of magnitude level as their actual values during close encounters. Given typical values for $a_*$ and $a_\cm$, we estimate this ratio to be $3\%$, which means a few percent of the angular momentum can be transferred between the stellar binary and its centre of mass orbit.

The bottom left panel of Figure~(\ref{fig:Ex}) shows the change in angular momentum of the SMBH-IMBH binary. Opposite to the middle panel of Figure~(\ref{fig:Ex}), we show that prograde orbits with $\mathbf{L}_{\cm,z}>0$ are more likely to lose angular momentum, while retrograde orbits with $\mathbf{L}_{\cm,z}<0$ are more likely to gain angular momentum. This is constrained by conservation of total angular momentum. However, due to the fact that HVBs in prograde orbits are more common than HVBs in retrograde orbits, the averaged effect of close encounters would make the SMBH-IMBH binary lose angular momentum.

Figure~(\ref{fig:progInnerFlip}) shows examples of
 \yw{simulations 
in which a strong interaction between the IMBH and the stellar binary 
results in a spin-flip of the latter and leads to an HVB.}  In each panel, from top to bottom, we display the energy and angular momentum exchange, the distance between the stellar binary and the IMBH at the time of ejection, and the total relative energy and angular momentum fluctuations, respectively. Since the z-component of the angular momentum of the black hole binary is positive in our simulations, positive $\mathbf{L}_{*,z}$($\mathbf{L}_{\cm,z}$) means prograde while negative $\mathbf{L}_{*,z}$($\mathbf{L}_{\cm,z}$) means retrograde. 

In the top left panel, the initial $\mathbf{L}_{\cm,z}(0)$ is positive while the initial $\mathbf{L}_{*,z}(0)$ is negative. Therefore, this is a prograde case with retrograde inner binary.  From the plot, we see that both the centre of mass orbit and the stellar binary orbit receive angular momentum from the IMBH, and the inner stellar orbit flips from retrograde $\mathbf{L}_{*,z} < 0$ to prograde $\mathbf{L}_{*,z} > 0$.

In the top right panel, the initial $\mathbf{L}_{\cm,z}(0)$ and $\mathbf{L}_{*,z}(0)$ are both positive; thus, this is a prograde case with prograde inner binary. From the upper-most subplot, we see that the centre of mass orbit gains angular momentum from the IMBH, while the stellar binary loses angular momentum and flips from prograde $\mathbf{L}_{*,z} > 0$ to retrograde $\mathbf{L}_{*,z} < 0$.

In the bottom left panel, the initial $\mathbf{L}_{\cm,z}(0)$ and $\mathbf{L}_{*,z}(0)$ are both negative; hence this is a retrograde case with retrograde inner binary. The upper-most subplot shows how both the centre of mass orbit and the stellar binary gain angular momentum from the IMBH, and $\mathbf{L}_{*,z}$ flips from retrograde $\mathbf{L}_{*,z} < 0$ to prograde $\mathbf{L}_{*,z} > 0$.

Last, in the bottom right panel, the initial $\mathbf{L}_{\cm,z}(0)$ is negative while the initial $\mathbf{L}_{*,z}(0)$ is positive. Therefore, this is a retrograde case with a prograde inner binary. In this case, the centre of mass orbit gains angular momentum from the IMBH, while the stellar binary orbit loses angular momentum. The stellar binary orbit remains prograde with $\mathbf{L}_{*,z} > 0$.

\subsection{Energy exchange}

The right two panels of Figure~(\ref{fig:Ex}) show the change in total energy before and after the close encounter. $E_b$ represents the energy of the stellar binary which is equal to:
\begin{equation}
E_* = -{G(m_{* 1}+m_{* 2})}/{2a_*}
\end{equation}
while $E_\cm$ is the energy corresponding to the centre of mass orbit of the stellar binary which is equal to:
\begin{equation}
E_\cm = -{G(m_{* 1}+m_{* 2}+m_{\bullet 1})}/{2a_\cm}
\end{equation}
As for the change in orbital energy, $\Delta E$ is calculated by comparing the initial and final energies.  As shown in the upper right panel of Figure~(\ref{fig:Ex}), $\Delta E_*/|E_*|$ cannot be larger than $1$. This is because $E_*$ is negative, and the stellar binary must become unbound to produce an HVB. If $\Delta E_*/|E_*|$ is larger than $1$, the energy of the stellar binary becomes positive leading to its disruption. {Both prograde orbits with $\mathbf{L}_{\cm,z}>0$ and retrograde orbits with $\mathbf{L}_{\cm,z}<0$ can gain or lose energy from interacting with the IMBH
, but retrograde orbits are more likely to lose more energy from the interaction, such that the stellar binary becomes more compact.} The middle right panel of Figure~(\ref{fig:Ex}) can be understood by noting that all HVBs escape hyperbolically from the SMBH, such that their centre of mass orbit must gain enough energy from the IMBH to make its total energy positive, which corresponds to the case $\Delta E_\cm/|E_\cm| > 1$. The bottom right panel of Figure~(\ref{fig:progInnerFlip}) shows that the energy change of the SMBH-IMBH binary $\Delta E_\bh$ is always negative, which implies that the SMBH-IMBH binary loses energy during the interaction with the stellar binaries.

\subsection{Properties of HVBs}
\begin{figure*}
\centering
\includegraphics[width=1\columnwidth]{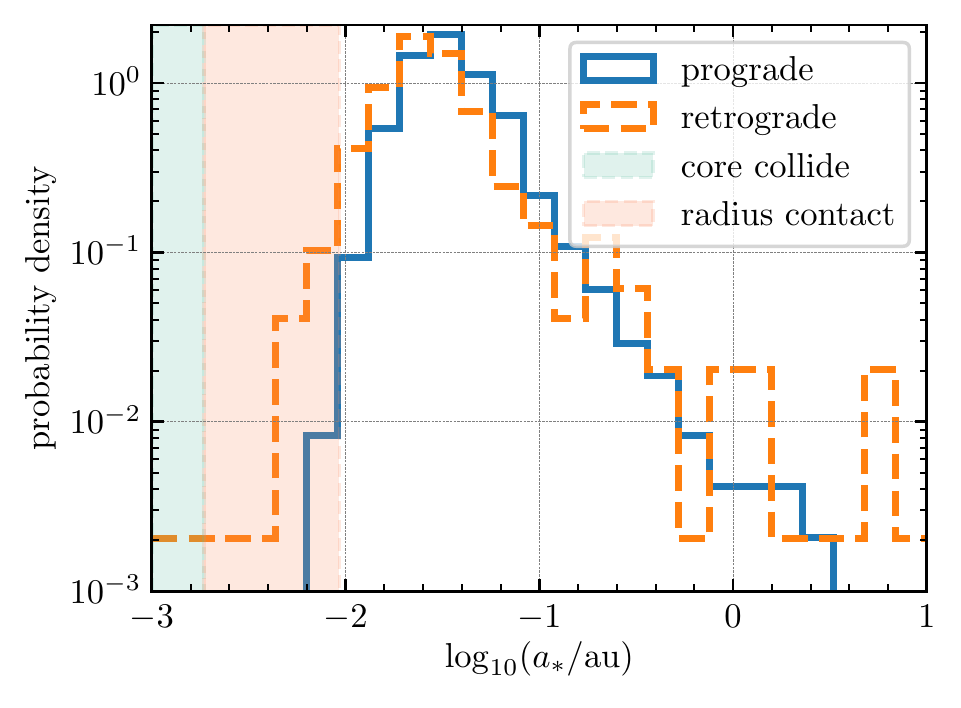}
\includegraphics[width=1\columnwidth]{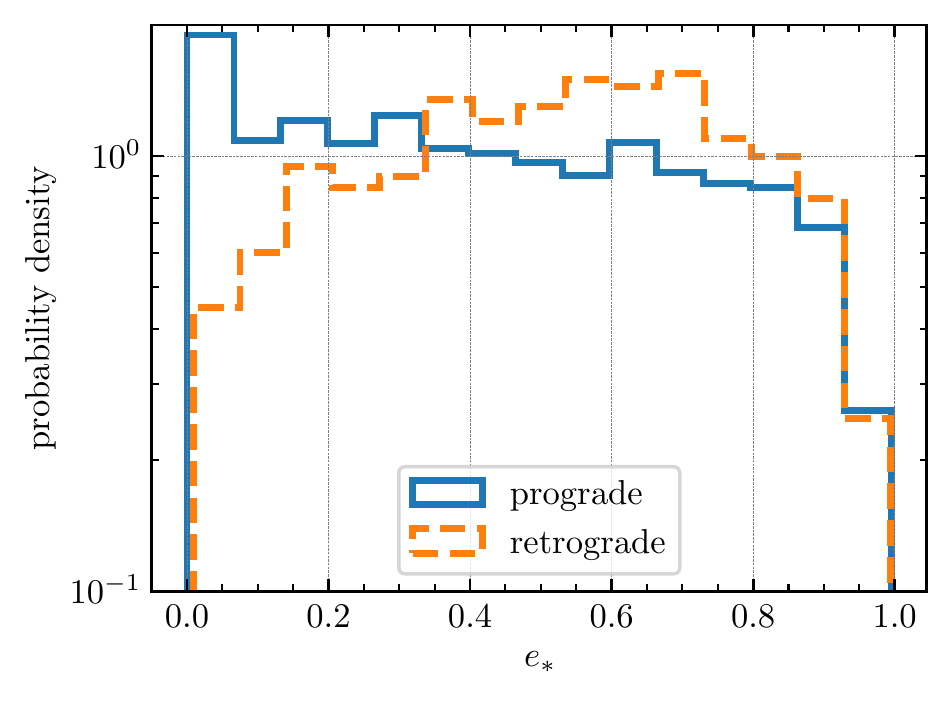}\\
\includegraphics[width=1\columnwidth]{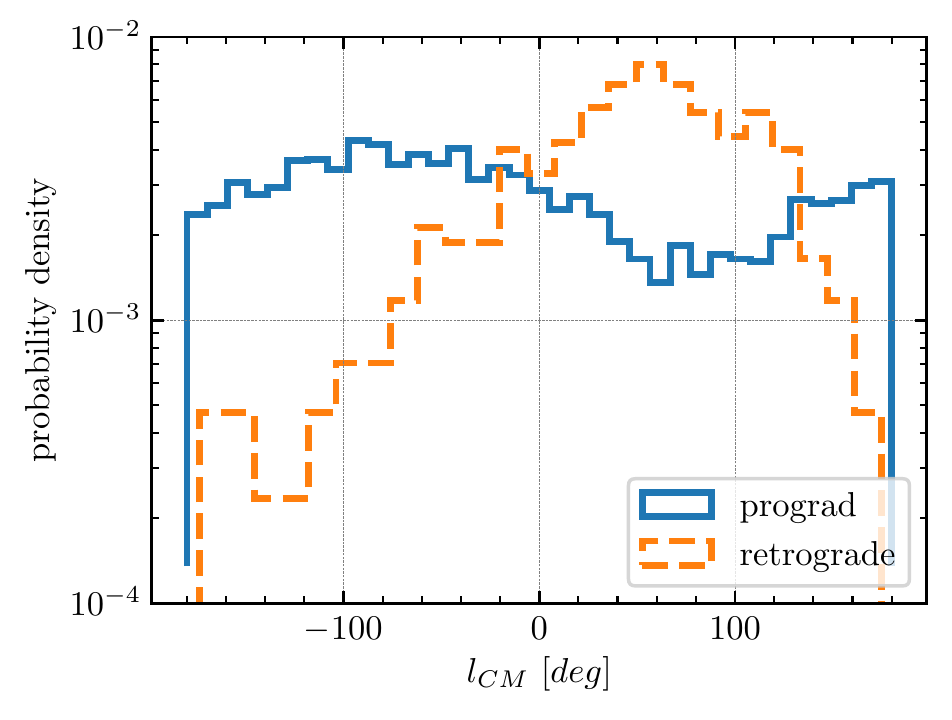}
\includegraphics[width=1\columnwidth]{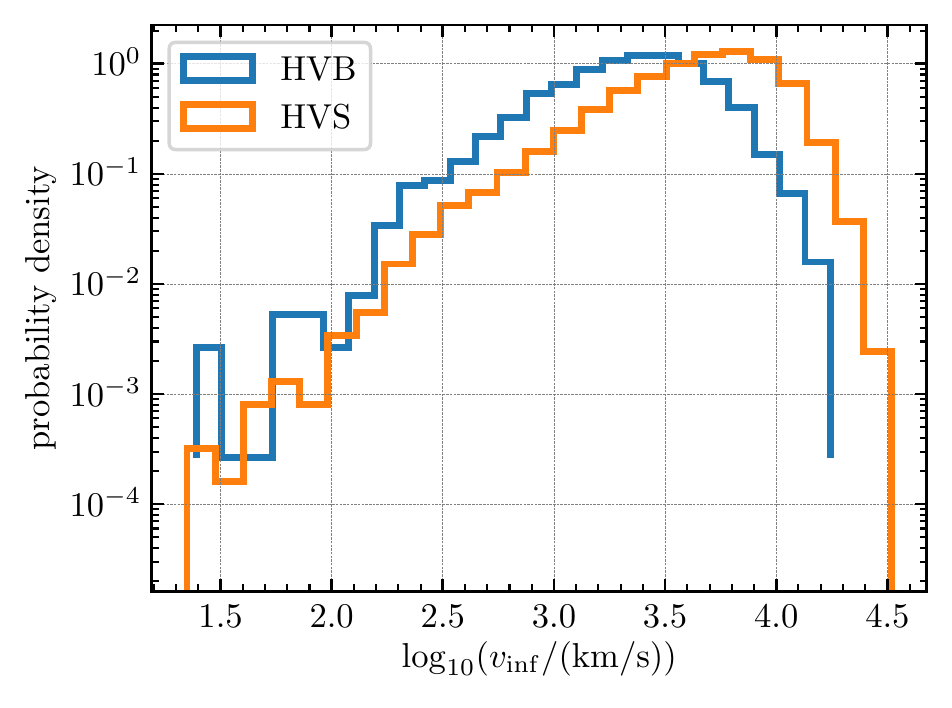}\\
\caption{Properties of the produced HVBs. The upper left panel shows the semi-major axis of the HVBs. HVBs from retrograde orbits are more compact than those from prograde orbits. The upper right panel shows the eccentricity of the HVBs. HVBs from retrograde orbits are more eccentric than those from prograde orbits. The bottom left panel shows the ejection direction of the HVBs. Finally, the bottom right panel shows the velocity distribution at infinity of the HVBs and HVSs.}
\label{fig:HVBpropt}
\end{figure*}

The coplanar configuration effectively ejects the stellar binary. In this section, we discuss the properties of these ejected stellar binaries, and in particular the statistics of the semi-major axis, eccentricity and ejection orientation post-ejection for both the prograde and retrograde cases. We also measure the velocity after ejection and estimate the velocity of the ejected binary at infinity \yw{without stellar evolution by} assuming a hyperbolic Keplerian orbit and integrate over an analytic form for the galactic potential \citep[e.g.,][]{2008ApJ...680..312K,2014ApJ...793..122K},

\begin{equation}
\Phi_G = \Phi_b + \Phi_d + \Phi_h
\end{equation}
with,
\begin{enumerate}{}{}
\item\textit{Bulge:}
\begin{equation}
\Phi_b(r)=-\frac{GM_b}{r+r_b}
\end{equation}
with $M_b=3.75\times 10^9 m_\odot$, $r_b=105$\,\pc\,  and $r$ in cylindrical coordinates,
\begin{equation}
r = \sqrt{\rho^2+z^2}
\end{equation}

\item\textit{Stellar disc:}
\begin{equation}\label{eq:stellarDsk}
\Phi_d(\rho,z)=-\frac{GM_d}{\sqrt{\rho^2+[a_d + (z^2+b_d^2)^{1/2} ]^2 }}
\end{equation}
with $M_d=6\times 10^{10} m_\odot$, $a_d=2.75$\,\kpc ~ and $b_d=300$\,\pc.

\item\textit{Dark-matter halo:}
\begin{equation}
\Phi_h(r)=-\frac{GM_h}{r}ln(1+r/r_h)
\end{equation}
with $M_h=1\times 10^{12} m_\odot$ and $r_h=20$\,\kpc.
\end{enumerate}

Figure~(\ref{fig:HVBpropt}) shows the probability density of HVBs as a function of $a_*$, $e_*$, $l_\cm$ and velocity at infinity $v_{inf}$. The upper left panel indicates that HVBs from retrograde orbits tend to be slightly more compact than HVBs from prograde orbits. The most probable range of the semi-major axes of HVBs is between $10^{-2}$ to $10^{-1}$ au. We also see some HVBs in the region of parameter space corresponding to contact of the stellar radii. This could potentially provide a mechanism for producing HVBs in a contact state, such as W UMa binaries, which could later merge to produce hypervelocity blue stragglers. The upper right panel shows the eccentricity distribution of the HVBs. \yw{As a consequence of the energy and angular momentum change in the stellar binaries shown in }the upper two panels of Figure~(\ref{fig:Ex}), the stellar binary eccentricity tends to increase after the ejection due to energy and angular momentum exchange with the IMBH. HVBs from retrograde orbits are more eccentric than those from prograde orbits. We note that, although we only considered main-sequence stellar binaries in this project, our results can be perspectively applied to other compact objects. These eccentric hypervelocity binaries, if composed of compact objects, could provide a source of intergalatic merger events due to gravitational radiation. Hence, these merger events would be detected off-centre of their origin galaxies. The last panel shows the velocity distribution of HVBs and HVSs at infinity. Limited by the tidal force from the IMBH, HVSs can get closer to the IMBH than HVBs.  Thus HVSs can get higher local velocity kicks from the IMBH. The typical tidal disruption radius of the stellar binary is given by Equation~(\ref{eq:rbt}) and the relationship between the local velocity kick $\Delta v_\cm$ and the distance of closest approach $r_{\min,2}$ is given by Equation~(\ref{eq:v-q}). Therefore, we estimate the maximum local kick velocity by assuming $r_{\min,2} = r_{bt}$ in Equation~(\ref{eq:v-q}). Rewriting the maximum local velocity kick,
\begin{equation}
\Delta v_{\max} = \bigg( \frac{m_b}{m_{\bullet 2}}\bigg)^{1/6}\bigg(\frac{Gm_{\bullet 2}}{r_{*,\min}}\bigg)^{1/2}\,,
\end{equation}
where $r_{*,\min}$ is the allowed smallest separation between the two components of the binary. In our simulations, this distance is given by the merger criterion in Section~(\ref{sec:criteria}).
\begin{figure}
\centering
\includegraphics[width=\columnwidth]{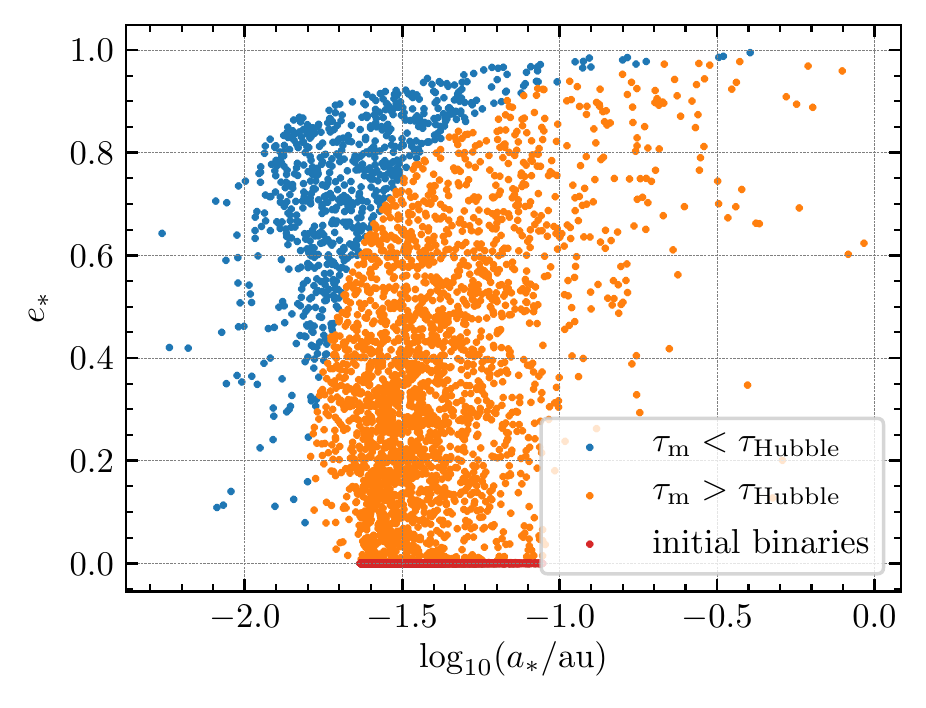}
\caption{The semi-major axis and eccentricity distributions of the ejected HVBs, along with their initial distributions. After the ejection, the eccentricities of the HVBs tends to increase, which could accelerate the gravitational wave radiation significantly if the binary components are compact objects. }
\label{fig:merger}
\end{figure}
Figure~(\ref{fig:merger}) shows the ejected binaries in the $a_{*}$-, $e_{*}$-parameter space. The merger time of isolated binaries due to gravitational wave radiation is given by \citep[e.g.,][]{1964PhRv..136.1224P},
\begin{equation}
\tau_m = \frac{3}{85}\frac{c^5a_*^4(1-e_*^2)^{7/2}}{G^3m_{*1}m_{*2}(m_{*1}+m_{*2})}\,.
\end{equation}
The red dots indicate the initial stellar binaries in the $e_*$-,$a_*$-parameter space, while the blue points indicate those ejected binaries for which the merger time is shorter than a Hubble time. The IMBH ejects the stellar binaries, pushing them into the merger regime where $\tau_m < \tau_{Hubble}$. Detailed investigation of other types of hypervelocity binaries will be performed in future works.

\section{Discussion and Summary} 

\subsection{Implications for the MW centre and other galaxies}
\label{discussion}
Our original set of 150~k simulations is drawn from a distribution of semimajor axes randomly sampled in the range $[0.2a_{\rm BH}, 3a_{\rm BH}]$, which is consistent with a density distribution $\rho(r)\propto r^{-2}$, i.e. an isothermal sphere. We can therefore consider the results as representative of the dynamical evolution of an IMBH into a cuspy stellar distribution, as predicted from theoretical models \citep{1977ApJ...216..883B} and observed in the MW centre, \citep[albeit flatter than $\rho(r)\propto r^{-2}$,][]{2018A&A...609A..28B} and other galaxies. Although a proper treatment of the problem should include self-consistently the orbital decay of the IMBH as well as interactions with unbound stars, we can still use our findings at face-value to get order of magnitude estimates. Out of 150k interactions, we find 88 HVB ejections, i.e. $\approx 0.06$\% of the total number of simulations. On its way to coalescence, an IMBH will interact with a mass in stars on the order of a few times its own mass. For example, \cite{2018MNRAS.476.4697M} find that an IMBH of $5\times 10^3\msun$ inspiralling onto SgrA$^*$ ejects about $4\times 10^4\msun$ in stars along the way. If we consider an interacting mass of stars of the order of $10~M_{\rm IMBH}$, and typical stellar masses of $1\msun$, we estimate about 50k interactions. If about 10\% of the interacting systems are compact binaries of the type simulated here, we then expect $\approx 3$ HVBs (or, in general, a few) ejected by an IMBH inspiralling onto SgrA$^*$.  There are at least 3 known binaries within the central 0.2 pc of the Galactic Centre \citep[e.g.][]{naoz18,ott99,rafelski07}, and it remains possible that many of the S-stars are actually unresolved binaries \citep[e.g.][]{naoz18}, although S0-2 which has the closest pericentre to Sgr A* has received relatively careful scrutiny in recent years but could technically still be a binary \citep{devin18}.

Additionally, note that, since the SMBH-IMBH binary is compact (as constrained by the observational data) with $v_{\bullet 2}$ being very large, the ejection velocity becomes very large (see Equation~\ref{eq:v-q}). The Galactic potential cannot efficiently slow down the ejected objects. The velocity distribution in the halo (i.e. say at 30~\kpc) looks almost the same as the one at infinity but with a lower limit truncation at 1000~km/s.

Upon ejection, HVBs can acquire a significant eccentricity. This can lead to excitation of tides leading to subsequent orbital shrinkage via tidal dissipation, or it can directly cause the gravitational wave merger timescale to become much shorter than a Hubble time (if the stellar binary is extremely compact). Either way, this is a promising avenue to produce a rejuvenated blue straggler like the B-type hypervelocity star (HVS) HE0437-5439, which, if ejected from the MW center \yw{or originated from the LMC \citep{2018arXiv180410197E},} has a travel time that is longer than its main sequence lifetime \citep{2005ApJ...622L..33B,2005ApJ...634L.181E}.  With that said, this result has recently been challenged, suggesting a Large Magellanic Cloud origin for this object \citep[e.g.][]{2018arXiv180705909I}.  However, such a high 3-D velocity could still be indicative of a massive BH binary, and certainly requires an anomalously compact and/or massive object such as a BH to generate the required acceleration.

Another interesting finding in our simulations is that stellar binaries can be significantly softened by close encounter with the IMBH leading to their eventual ejection. The top left panel of Figure~(\ref{fig:HVBpropt}) shows, in fact, that there is an $\approx 0.3$\% probability of ejecting a stellar binary with $a_*>1$AU. Although this probability is small, it is not null; we therefore conclude that it is not impossible to produce soft HVBs similar to SDSS J1211+1437 \citep{2016ApJ...821L..13N} via dynamical ejection. 

Looking beyond the MW, the ejection of HVBs appears to be a promising route to produce exotic phenomena in the intergalactic space. As shown in Figure~(\ref{fig:merger}), the merger timescale of such binaries can be shorter than a Hubble time. If an HVB travels at $\approx 1000$~km~s$^{-1}$ for a Gyr before merging, the merger will occur at a distance $\approx 1$~\Mpc\, from the ejection site. Although we consider regular stars in this first investigation, this opens many interesting prospects for detecting mergers of compact objects in intergalactic space. These might include Type Ia Supernovae produced by merging white dwarf binaries \citep{2014ARA&A..52..107M} or Gamma-Ray Bursts with associated GW detection produced by merging neutron star binaries \citep{2017ApJ...848L..12A}. For intergalactic merging black hole binaries, however, since they are expected to occur without electromagnetic counterparts, it will be more difficult to discern whether the merger occurred inside or outside a specific galaxy. Note that, although HVB ejection probabilities are small from SMBH-IMBH binaries embedded in a spherical cluster of stars, they are substantially larger if the system is in a disk-like configuration. This might happen if the IMBH is inspiralling into an AGN disk that is fragmenting and producing stars and binaries \citep{2018arXiv180702859S}. In our set of 300\,k coplanar simulations we find HVB production fractions of about 1\%, 20 times higher than in the spherical cluster case. Moreover, \cite{2018MNRAS.475.4595W} found that the HVB ejection fractions are substantially larger for SMBHBs composed of SMBHs with comparable masses. We will explore the comsic rate of such events in future work.

\subsection{Summary}
In this work we have studied hypervelocity binaries ejected by massive black hole binaries, as a function of the orbital inclinations and interaction parameters as well as the properties of the ejected stellar binaries. We have performed high-precision N-body simulations of the orbital evolution of stellar binary-SMBH binary systems, and analyzed the outcomes. Our main conclusions can be summarized as follows. 

Our simulations show that hypervelocity binaries are efficiently produced in the inner stable region of the four-body system.  This is defined in our simulations as a configuration in which the stellar binary and the SMBH form a stable hierarchical triple, while the SMBH-stellar binary-IMBH forms an unstable triple. In such a system, the inner hierarchical triple guarantees that the stellar binary is not tidally disrupted by the central SMBH; at the same time, the unstable outer triple creates an increased probability for stellar binaries to interact directly with the IMBH. In this region of parameter space, the stellar binaries efficiently encounter the IMBH with the SMBH acting as a remote perturber, which leads to stellar binary ejections.

Hypervelocity binaries tend to be ejected at low relative inclinations of the three main orbits involved in the system. This is because, at low inclinations, the Kozai-Lidov oscillations that create most of the mergers and TDEs are suppressed. Therefore, a large number of stellar binaries can survive long enough to undergo close interactions with the IMBH and eventually be ejected. Note also that HVBs are produced almost exclusively in coplanar prograde orbits, rarely occurring in retrograde ones, which can be simply understood by the much smaller interaction cross section with the IMBH in the latter case. For prograde orbits, the relative velocity of the IMBH and the binary centre of mass can be arbitrarily small, greatly enhancing the interaction cross section; for retrograde orbits, the typical relative velocity at interaction is instead $2\,v_{\rm IMBH}$. A promising channel to form low inclination prograde orbits would be an IMBH residing in the disk of an active galactic nucleus \citep{2018arXiv180702859S}.

Ejected hypervelocity binaries can acquire a high eccentricity from the close encounter with the IMBH. Therefore, the stars in these high eccentricity compact hypervelocity binaries will become more prone to interact strongly with their companions. Furthermore, these eccentric compact binaries are likely to merge within a Hubble time due to gravitational radiation losses. Thus, if composed of compact objects (either neutron stars or black holes), HVBs produced from a close interaction with the IMBH in an SMBH-IMBH binary could be regarded as a promising extragalactic merger resource for ground based interferometers. We plan to explore this formation channel of GW sources in future work.

Our results show that hypervelocity binaries are efficiently produced by SMBH-SMBH/SMBH-IMBH binaries with large mass ratios \citep[e.g.,][]{2018MNRAS.475.4595W} and low inclinations. The ejected centre of mass velocity of the stellar binary is proportional to the mass of the secondary massive black hole to the one-third power ($\sim m_\imbh^{1/3}$). Since HVBs with extreme velocities $\gtrsim$~1000~km/s can only be produced by super-massive black hole binaries, the discovery of even a single  hypervelocity binary in the Milky Way with such a high speed could be regarded as a smoking gun for the presence of an intermediate-mass black hole in the Galactic Centre, with a mass in the range constrained by current observational limits.

\section*{Acknowledgments}
Yihan Wang thanks Mario Spera for introducing the {\tt ARCHAIN} algorithm, Bin Liu and Dong Lai for helpful discussion.

\label{lastpage}
\bibliography{ref}

\begin{thebibliography}{}
\makeatletter
\relax
\def\mn@urlcharsother{\let\do\@makeother \do\$\do\&\do\#\do\^\do\_\do\%\do\~}
\def\mn@doi{\begingroup\mn@urlcharsother \@ifnextchar [ {\mn@doi@}
  {\mn@doi@[]}}
\def\mn@doi@[#1]#2{\def\@tempa{#1}\ifx\@tempa\@empty \href
  {http://dx.doi.org/#2} {doi:#2}\else \href {http://dx.doi.org/#2} {#1}\fi
  \endgroup}
\def\mn@eprint#1#2{\mn@eprint@#1:#2::\@nil}
\def\mn@eprint@arXiv#1{\href {http://arxiv.org/abs/#1} {{\tt arXiv:#1}}}
\def\mn@eprint@dblp#1{\href {http://dblp.uni-trier.de/rec/bibtex/#1.xml}
  {dblp:#1}}
\def\mn@eprint@#1:#2:#3:#4\@nil{\def\@tempa {#1}\def\@tempb {#2}\def\@tempc
  {#3}\ifx \@tempc \@empty \let \@tempc \@tempb \let \@tempb \@tempa \fi \ifx
  \@tempb \@empty \def\@tempb {arXiv}\fi \@ifundefined
  {mn@eprint@\@tempb}{\@tempb:\@tempc}{\expandafter \expandafter \csname
  mn@eprint@\@tempb\endcsname \expandafter{\@tempc}}}

\bibitem[\protect\citeauthoryear{{Abadi}, {Navarro}  \& {Steinmetz}}{{Abadi}
  et~al.}{2009}]{2009ApJ...691L..63A}
{Abadi} M.~G.,  {Navarro} J.~F.,   {Steinmetz} M.,  2009, \mn@doi [\apjl]
  {10.1088/0004-637X/691/2/L63}, \href
  {http://adsabs.harvard.edu/abs/2009ApJ...691L..63A} {691, L63}

\bibitem[\protect\citeauthoryear{{Abbott} et~al.,}{{Abbott}
  et~al.}{2017}]{2017ApJ...848L..12A}
{Abbott} B.~P.,  et~al., 2017, \mn@doi [\apjl] {10.3847/2041-8213/aa91c9},
  \href {http://adsabs.harvard.edu/abs/2017ApJ...848L..12A} {848, L12}

\bibitem[\protect\citeauthoryear{{Antognini}, {Shappee}, {Thompson}  \&
  {Amaro-Seoane}}{{Antognini} et~al.}{2014}]{2014MNRAS.439.1079A}
{Antognini} J.~M.,  {Shappee} B.~J.,  {Thompson} T.~A.,   {Amaro-Seoane} P.,
  2014, \mn@doi [\mnras] {10.1093/mnras/stu039}, \href
  {http://adsabs.harvard.edu/abs/2014MNRAS.439.1079A} {439, 1079}

\bibitem[\protect\citeauthoryear{{Bahcall} \& {Wolf}}{{Bahcall} \&
  {Wolf}}{1977}]{1977ApJ...216..883B}
{Bahcall} J.~N.,  {Wolf} R.~A.,  1977, \mn@doi [\apj] {10.1086/155534}, \href
  {http://adsabs.harvard.edu/abs/1977ApJ...216..883B} {216, 883}

\bibitem[\protect\citeauthoryear{{Baumgardt}, {Gualandris}  \& {Portegies
  Zwart}}{{Baumgardt} et~al.}{2006}]{2006MNRAS.372..174B}
{Baumgardt} H.,  {Gualandris} A.,   {Portegies Zwart} S.,  2006, \mn@doi
  [\mnras] {10.1111/j.1365-2966.2006.10818.x}, \href
  {http://adsabs.harvard.edu/abs/2006MNRAS.372..174B} {372, 174}

\bibitem[\protect\citeauthoryear{{Baumgardt}, {Amaro-Seoane}  \&
  {Sch{\"o}del}}{{Baumgardt} et~al.}{2018}]{2018A&A...609A..28B}
{Baumgardt} H.,  {Amaro-Seoane} P.,   {Sch{\"o}del} R.,  2018, \mn@doi [\aap]
  {10.1051/0004-6361/201730462}, \href
  {http://adsabs.harvard.edu/abs/2018A\%26A...609A..28B} {609, A28}

\bibitem[\protect\citeauthoryear{{Blaauw}}{{Blaauw}}{1961}]{1961BAN....15..265B}
{Blaauw} A.,  1961, \bain, \href
  {http://adsabs.harvard.edu/abs/1961BAN....15..265B} {15, 265}

\bibitem[\protect\citeauthoryear{{Boubert}, {Guillochon}, {Hawkins},
  {Ginsburg}, {Evans}  \& {Strader}}{{Boubert}
  et~al.}{2018}]{2018MNRAS.479.2789B}
{Boubert} D.,  {Guillochon} J.,  {Hawkins} K.,  {Ginsburg} I.,  {Evans} N.~W.,
   {Strader} J.,  2018, \mn@doi [\mnras] {10.1093/mnras/sty1601}, \href
  {http://adsabs.harvard.edu/abs/2018MNRAS.479.2789B} {479, 2789}

\bibitem[\protect\citeauthoryear{{Bromley}, {Kenyon}, {Brown}  \&
  {Geller}}{{Bromley} et~al.}{2009}]{2009ApJ...706..925B}
{Bromley} B.~C.,  {Kenyon} S.~J.,  {Brown} W.~R.,   {Geller} M.~J.,  2009,
  \mn@doi [\apj] {10.1088/0004-637X/706/2/925}, \href
  {http://adsabs.harvard.edu/abs/2009ApJ...706..925B} {706, 925}

\bibitem[\protect\citeauthoryear{{Brown}, {Geller}, {Kenyon}  \&
  {Kurtz}}{{Brown} et~al.}{2005}]{2005ApJ...622L..33B}
{Brown} W.~R.,  {Geller} M.~J.,  {Kenyon} S.~J.,   {Kurtz} M.~J.,  2005,
  \mn@doi [\apjl] {10.1086/429378}, \href
  {http://adsabs.harvard.edu/abs/2005ApJ...622L..33B} {622, L33}

\bibitem[\protect\citeauthoryear{{Brown}, {Geller}, {Kenyon}  \&
  {Kurtz}}{{Brown} et~al.}{2006}]{2006ApJ...647..303B}
{Brown} W.~R.,  {Geller} M.~J.,  {Kenyon} S.~J.,   {Kurtz} M.~J.,  2006,
  \mn@doi [\apj] {10.1086/505165}, \href
  {http://adsabs.harvard.edu/abs/2006ApJ...647..303B} {647, 303}

\bibitem[\protect\citeauthoryear{{Brown}, {Anderson}, {Gnedin}, {Bond},
  {Geller}, {Kenyon}  \& {Livio}}{{Brown} et~al.}{2010}]{2010ApJ...719L..23B}
{Brown} W.~R.,  {Anderson} J.,  {Gnedin} O.~Y.,  {Bond} H.~E.,  {Geller} M.~J.,
   {Kenyon} S.~J.,   {Livio} M.,  2010, \mn@doi [\apjl]
  {10.1088/2041-8205/719/1/L23}, \href
  {http://adsabs.harvard.edu/abs/2010ApJ...719L..23B} {719, L23}

\bibitem[\protect\citeauthoryear{{Brown}, {Geller}  \& {Kenyon}}{{Brown}
  et~al.}{2014}]{2014ApJ...787...89B}
{Brown} W.~R.,  {Geller} M.~J.,   {Kenyon} S.~J.,  2014, \mn@doi [\apj]
  {10.1088/0004-637X/787/1/89}, \href
  {http://adsabs.harvard.edu/abs/2014ApJ...787...89B} {787, 89}

\bibitem[\protect\citeauthoryear{{Brown}, {Anderson}, {Gnedin}, {Bond},
  {Geller}  \& {Kenyon}}{{Brown} et~al.}{2015}]{2015ApJ...804...49B}
{Brown} W.~R.,  {Anderson} J.,  {Gnedin} O.~Y.,  {Bond} H.~E.,  {Geller} M.~J.,
    {Kenyon} S.~J.,  2015, \mn@doi [\apj] {10.1088/0004-637X/804/1/49}, \href
  {http://adsabs.harvard.edu/abs/2015ApJ...804...49B} {804, 49}

\bibitem[\protect\citeauthoryear{{Chen}, {Madau}, {Sesana}  \& {Liu}}{{Chen}
  et~al.}{2009}]{2009ApJ...697L.149C}
{Chen} X.,  {Madau} P.,  {Sesana} A.,   {Liu} F.~K.,  2009, \mn@doi [\apjl]
  {10.1088/0004-637X/697/2/L149}, \href
  {http://adsabs.harvard.edu/abs/2009ApJ...697L.149C} {697, L149}

\bibitem[\protect\citeauthoryear{{Chu} et~al.,}{{Chu} et~al.}{2018}]{devin18}
{Chu} D.~S.,  et~al., 2018, \mn@doi [\apj] {10.3847/1538-4357/aaa3eb}, \href
  {http://adsabs.harvard.edu/abs/2018ApJ...854...12C} {854, 12}

\bibitem[\protect\citeauthoryear{{Colgate}}{{Colgate}}{1970}]{1970Natur.225..247C}
{Colgate} S.~A.,  1970, \mn@doi [\nat] {10.1038/225247a0}, \href
  {http://adsabs.harvard.edu/abs/1970Natur.225..247C} {225, 247}

\bibitem[\protect\citeauthoryear{{Edelmann}, {Napiwotzki}, {Heber},
  {Christlieb}  \& {Reimers}}{{Edelmann} et~al.}{2005}]{2005ApJ...634L.181E}
{Edelmann} H.,  {Napiwotzki} R.,  {Heber} U.,  {Christlieb} N.,   {Reimers} D.,
   2005, \mn@doi [\apjl] {10.1086/498940}, \href
  {http://adsabs.harvard.edu/abs/2005ApJ...634L.181E} {634, L181}

\bibitem[\protect\citeauthoryear{{Erkal}, {Boubert}, {Gualandris}, {Evans}  \&
  {Antonini}}{{Erkal} et~al.}{2018}]{2018arXiv180410197E}
{Erkal} D.,  {Boubert} D.,  {Gualandris} A.,  {Evans} N.~W.,   {Antonini} F.,
  2018, preprint, \href {http://adsabs.harvard.edu/abs/2018arXiv180410197E} {}
  (\mn@eprint {arXiv} {1804.10197})

\bibitem[\protect\citeauthoryear{{Fragione}}{{Fragione}}{2018}]{2018MNRAS.479.2615F}
{Fragione} G.,  2018, \mn@doi [\mnras] {10.1093/mnras/sty1593}, \href
  {http://adsabs.harvard.edu/abs/2018MNRAS.479.2615F} {479, 2615}

\bibitem[\protect\citeauthoryear{{Fragione} \& {Gualandris}}{{Fragione} \&
  {Gualandris}}{2018}]{2018MNRAS.475.4986F}
{Fragione} G.,  {Gualandris} A.,  2018, \mn@doi [\mnras]
  {10.1093/mnras/sty145}, \href
  {http://adsabs.harvard.edu/abs/2018MNRAS.475.4986F} {475, 4986}

\bibitem[\protect\citeauthoryear{{Fragione}, {Capuzzo-Dolcetta}  \&
  {Kroupa}}{{Fragione} et~al.}{2017}]{2017MNRAS.467..451F}
{Fragione} G.,  {Capuzzo-Dolcetta} R.,   {Kroupa} P.,  2017, \mn@doi [\mnras]
  {10.1093/mnras/stx106}, \href
  {http://adsabs.harvard.edu/abs/2017MNRAS.467..451F} {467, 451}

\bibitem[\protect\citeauthoryear{{Fragione}, {Leigh}, {Ginsburg}  \&
  {Kocsis}}{{Fragione} et~al.}{2018a}]{2018arXiv180608385F}
{Fragione} G.,  {Leigh} N.,  {Ginsburg} I.,   {Kocsis} B.,  2018a, preprint,
  \href {http://adsabs.harvard.edu/abs/2018arXiv180608385F} {} (\mn@eprint
  {arXiv} {1806.08385})

\bibitem[\protect\citeauthoryear{{Fragione}, {Ginsburg}  \&
  {Kocsis}}{{Fragione} et~al.}{2018b}]{2018ApJ...856...92F}
{Fragione} G.,  {Ginsburg} I.,   {Kocsis} B.,  2018b, \mn@doi [\apj]
  {10.3847/1538-4357/aab368}, \href
  {http://adsabs.harvard.edu/abs/2018ApJ...856...92F} {856, 92}

\bibitem[\protect\citeauthoryear{{Grishin}, {Perets}  \& {Fragione}}{{Grishin}
  et~al.}{2018}]{2018arXiv180802030G}
{Grishin} E.,  {Perets} H.~B.,   {Fragione} G.,  2018, preprint, \href
  {http://adsabs.harvard.edu/abs/2018arXiv180802030G} {} (\mn@eprint {arXiv}
  {1808.02030})

\bibitem[\protect\citeauthoryear{{Gualandris} \& {Merritt}}{{Gualandris} \&
  {Merritt}}{2009}]{2009ApJ...705..361G}
{Gualandris} A.,  {Merritt} D.,  2009, \mn@doi [\apj]
  {10.1088/0004-637X/705/1/361}, \href
  {http://adsabs.harvard.edu/abs/2009ApJ...705..361G} {705, 361}

\bibitem[\protect\citeauthoryear{{Hansen}, {Kawaler}  \& {Trimble}}{{Hansen}
  et~al.}{2004}]{2004sipp.book.....H}
{Hansen} C.~J.,  {Kawaler} S.~D.,   {Trimble} V.,  2004, {Stellar interiors :
  physical principles, structure, and evolution}

\bibitem[\protect\citeauthoryear{{Hills}}{{Hills}}{1988}]{1988Natur.331..687H}
{Hills} J.~G.,  1988, \mn@doi [\nat] {10.1038/331687a0}, \href
  {http://adsabs.harvard.edu/abs/1988Natur.331..687H} {331, 687}

\bibitem[\protect\citeauthoryear{{Irrgang}, {Kreuzer}  \& {Heber}}{{Irrgang}
  et~al.}{2018}]{2018arXiv180705909I}
{Irrgang} A.,  {Kreuzer} S.,   {Heber} U.,  2018, preprint, \href
  {http://adsabs.harvard.edu/abs/2018arXiv180705909I} {} (\mn@eprint {arXiv}
  {1807.05909})

\bibitem[\protect\citeauthoryear{{Kenyon}, {Bromley}, {Geller}  \&
  {Brown}}{{Kenyon} et~al.}{2008}]{2008ApJ...680..312K}
{Kenyon} S.~J.,  {Bromley} B.~C.,  {Geller} M.~J.,   {Brown} W.~R.,  2008,
  \mn@doi [\apj] {10.1086/587738}, \href
  {http://adsabs.harvard.edu/abs/2008ApJ...680..312K} {680, 312}

\bibitem[\protect\citeauthoryear{{Kenyon}, {Bromley}, {Brown}  \&
  {Geller}}{{Kenyon} et~al.}{2014}]{2014ApJ...793..122K}
{Kenyon} S.~J.,  {Bromley} B.~C.,  {Brown} W.~R.,   {Geller} M.~J.,  2014,
  \mn@doi [\apj] {10.1088/0004-637X/793/2/122}, \href
  {http://adsabs.harvard.edu/abs/2014ApJ...793..122K} {793, 122}

\bibitem[\protect\citeauthoryear{{Kiseleva}, {Aarseth}, {Eggleton}  \& {de La
  Fuente Marcos}}{{Kiseleva} et~al.}{1996}]{1996ASPC...90..433K}
{Kiseleva} L.~G.,  {Aarseth} S.~J.,  {Eggleton} P.~P.,   {de La Fuente Marcos}
  R.,  1996, in {Milone} E.~F.,  {Mermilliod} J.-C.,  eds,  Astronomical
  Society of the Pacific Conference Series Vol. 90, The Origins, Evolution, and
  Destinies of Binary Stars in Clusters. p.~433

\bibitem[\protect\citeauthoryear{{Liu}, {Wang}  \& {Yuan}}{{Liu}
  et~al.}{2017}]{2017MNRAS.466.3376L}
{Liu} B.,  {Wang} Y.-H.,   {Yuan} Y.-F.,  2017, \mn@doi [\mnras]
  {10.1093/mnras/stw3300}, \href
  {http://adsabs.harvard.edu/abs/2017MNRAS.466.3376L} {466, 3376}

\bibitem[\protect\citeauthoryear{{Lu}, {Yu}  \& {Lin}}{{Lu}
  et~al.}{2007}]{2007ApJ...666L..89L}
{Lu} Y.,  {Yu} Q.,   {Lin} D.~N.~C.,  2007, \mn@doi [\apjl] {10.1086/521708},
  \href {http://adsabs.harvard.edu/abs/2007ApJ...666L..89L} {666, L89}

\bibitem[\protect\citeauthoryear{{Maoz}, {Mannucci}  \& {Nelemans}}{{Maoz}
  et~al.}{2014}]{2014ARA&A..52..107M}
{Maoz} D.,  {Mannucci} F.,   {Nelemans} G.,  2014, \mn@doi [\araa]
  {10.1146/annurev-astro-082812-141031}, \href
  {http://adsabs.harvard.edu/abs/2014ARA\%26A..52..107M} {52, 107}

\bibitem[\protect\citeauthoryear{{Marchetti}, {Contigiani}, {Rossi}, {Albert},
  {Brown}  \& {Sesana}}{{Marchetti} et~al.}{2018}]{2018MNRAS.476.4697M}
{Marchetti} T.,  {Contigiani} O.,  {Rossi} E.~M.,  {Albert} J.~G.,  {Brown}
  A.~G.~A.,   {Sesana} A.,  2018, \mn@doi [\mnras] {10.1093/mnras/sty579},
  \href {http://adsabs.harvard.edu/abs/2018MNRAS.476.4697M} {476, 4697}

\bibitem[\protect\citeauthoryear{{Mardling} \& {Aarseth}}{{Mardling} \&
  {Aarseth}}{2001}]{2001MNRAS.321..398M}
{Mardling} R.~A.,  {Aarseth} S.~J.,  2001, \mn@doi [\mnras]
  {10.1046/j.1365-8711.2001.03974.x}, \href
  {http://adsabs.harvard.edu/abs/2001MNRAS.321..398M} {321, 398}

\bibitem[\protect\citeauthoryear{{McKernan}, {Ford}, {Lyra}  \&
  {Perets}}{{McKernan} et~al.}{2012}]{2012MNRAS.425..460M}
{McKernan} B.,  {Ford} K.~E.~S.,  {Lyra} W.,   {Perets} H.~B.,  2012, \mn@doi
  [\mnras] {10.1111/j.1365-2966.2012.21486.x}, \href
  {http://adsabs.harvard.edu/abs/2012MNRAS.425..460M} {425, 460}

\bibitem[\protect\citeauthoryear{{Mikkola} \& {Merritt}}{{Mikkola} \&
  {Merritt}}{2008}]{2008AJ....135.2398M}
{Mikkola} S.,  {Merritt} D.,  2008, \mn@doi [\aj]
  {10.1088/0004-6256/135/6/2398}, \href
  {http://adsabs.harvard.edu/abs/2008AJ....135.2398M} {135, 2398}

\bibitem[\protect\citeauthoryear{{Naoz}, {Ghez}, {Hees}, {Do}, {Witzel}  \&
  {Lu}}{{Naoz} et~al.}{2018}]{naoz18}
{Naoz} S.,  {Ghez} A.~M.,  {Hees} A.,  {Do} T.,  {Witzel} G.,   {Lu} J.~R.,
  2018, \mn@doi [\apjl] {10.3847/2041-8213/aaa6bf}, \href
  {http://adsabs.harvard.edu/abs/2018ApJ...853L..24N} {853, L24}

\bibitem[\protect\citeauthoryear{{N{\'e}meth}, {Ziegerer}, {Irrgang}, {Geier},
  {F{\"u}rst}, {Kupfer}  \& {Heber}}{{N{\'e}meth}
  et~al.}{2016}]{2016ApJ...821L..13N}
{N{\'e}meth} P.,  {Ziegerer} E.,  {Irrgang} A.,  {Geier} S.,  {F{\"u}rst} F.,
  {Kupfer} T.,   {Heber} U.,  2016, \mn@doi [\apjl]
  {10.3847/2041-8205/821/1/L13}, \href
  {http://adsabs.harvard.edu/abs/2016ApJ...821L..13N} {821, L13}

\bibitem[\protect\citeauthoryear{{Ott}, {Eckart}  \& {Genzel}}{{Ott}
  et~al.}{1999}]{ott99}
{Ott} T.,  {Eckart} A.,   {Genzel} R.,  1999, \mn@doi [\apj] {10.1086/307712},
  \href {http://adsabs.harvard.edu/abs/1999ApJ...523..248O} {523, 248}

\bibitem[\protect\citeauthoryear{{Perets}}{{Perets}}{2009}]{2009ApJ...698.1330P}
{Perets} H.~B.,  2009, \mn@doi [\apj] {10.1088/0004-637X/698/2/1330}, \href
  {http://adsabs.harvard.edu/abs/2009ApJ...698.1330P} {698, 1330}

\bibitem[\protect\citeauthoryear{{Peters}}{{Peters}}{1964}]{1964PhRv..136.1224P}
{Peters} P.~C.,  1964, \mn@doi [Physical Review] {10.1103/PhysRev.136.B1224},
  \href {http://adsabs.harvard.edu/abs/1964PhRv..136.1224P} {136, 1224}

\bibitem[\protect\citeauthoryear{{Portegies Zwart}, {Baumgardt}, {McMillan},
  {Makino}, {Hut}  \& {Ebisuzaki}}{{Portegies Zwart}
  et~al.}{2006}]{2006ApJ...641..319P}
{Portegies Zwart} S.~F.,  {Baumgardt} H.,  {McMillan} S.~L.~W.,  {Makino} J.,
  {Hut} P.,   {Ebisuzaki} T.,  2006, \mn@doi [\apj] {10.1086/500361}, \href
  {http://adsabs.harvard.edu/abs/2006ApJ...641..319P} {641, 319}

\bibitem[\protect\citeauthoryear{{Quinlan}}{{Quinlan}}{1996}]{1996NewA....1...35Q}
{Quinlan} G.~D.,  1996, \mn@doi [\na] {10.1016/S1384-1076(96)00003-6}, \href
  {http://adsabs.harvard.edu/abs/1996NewA....1...35Q} {1, 35}

\bibitem[\protect\citeauthoryear{{Rafelski}, {Ghez}, {Hornstein}, {Lu}  \&
  {Morris}}{{Rafelski} et~al.}{2007}]{rafelski07}
{Rafelski} M.,  {Ghez} A.~M.,  {Hornstein} S.~D.,  {Lu} J.~R.,   {Morris} M.,
  2007, \mn@doi [\apj] {10.1086/512062}, \href
  {http://adsabs.harvard.edu/abs/2007ApJ...659.1241R} {659, 1241}

\bibitem[\protect\citeauthoryear{{Secunda}, {Bellovary}, {Mac Low}, {Ford},
  {McKernan}, {Leigh}  \& {Lyra}}{{Secunda} et~al.}{2018}]{2018arXiv180702859S}
{Secunda} A.,  {Bellovary} J.,  {Mac Low} M.-M.,  {Ford} K.~E.~S.,  {McKernan}
  B.,  {Leigh} N.,   {Lyra} W.,  2018, preprint, \href
  {http://adsabs.harvard.edu/abs/2018arXiv180702859S} {} (\mn@eprint {arXiv}
  {1807.02859})

\bibitem[\protect\citeauthoryear{{Sesana}, {Haardt}  \& {Madau}}{{Sesana}
  et~al.}{2006}]{2006ApJ...651..392S}
{Sesana} A.,  {Haardt} F.,   {Madau} P.,  2006, \mn@doi [\apj]
  {10.1086/507596}, \href {http://adsabs.harvard.edu/abs/2006ApJ...651..392S}
  {651, 392}

\bibitem[\protect\citeauthoryear{{Sesana}, {Haardt}  \& {Madau}}{{Sesana}
  et~al.}{2007}]{2007MNRAS.379L..45S}
{Sesana} A.,  {Haardt} F.,   {Madau} P.,  2007, \mn@doi [\mnras]
  {10.1111/j.1745-3933.2007.00331.x}, \href
  {http://adsabs.harvard.edu/abs/2007MNRAS.379L..45S} {379, L45}

\bibitem[\protect\citeauthoryear{{Sesana}, {Madau}  \& {Haardt}}{{Sesana}
  et~al.}{2009}]{2009MNRAS.392L..31S}
{Sesana} A.,  {Madau} P.,   {Haardt} F.,  2009, \mn@doi [\mnras]
  {10.1111/j.1745-3933.2008.00578.x}, \href
  {http://adsabs.harvard.edu/abs/2009MNRAS.392L..31S} {392, L31}

\bibitem[\protect\citeauthoryear{{Stoer}}{{Stoer}}{1972}]{1972BAAS....4T.422S}
{Stoer} J.,  1972, in Bulletin of the American Astronomical Society. pp
  422--423

\bibitem[\protect\citeauthoryear{{Wang}, {Leigh}, {Yuan}  \& {Perna}}{{Wang}
  et~al.}{2018}]{2018MNRAS.475.4595W}
{Wang} Y.-H.,  {Leigh} N.,  {Yuan} Y.-F.,   {Perna} R.,  2018, \mn@doi [\mnras]
  {10.1093/mnras/sty107}, \href
  {http://adsabs.harvard.edu/abs/2018MNRAS.475.4595W} {475, 4595}

\bibitem[\protect\citeauthoryear{{Yu} \& {Tremaine}}{{Yu} \&
  {Tremaine}}{2003}]{2003ApJ...599.1129Y}
{Yu} Q.,  {Tremaine} S.,  2003, \mn@doi [\apj] {10.1086/379546}, \href
  {http://adsabs.harvard.edu/abs/2003ApJ...599.1129Y} {599, 1129}

\bibitem[\protect\citeauthoryear{{Zheng} et~al.,}{{Zheng}
  et~al.}{2014}]{2014ApJ...785L..23Z}
{Zheng} Z.,  et~al., 2014, \mn@doi [\apjl] {10.1088/2041-8205/785/2/L23}, \href
  {http://adsabs.harvard.edu/abs/2014ApJ...785L..23Z} {785, L23}

\makeatother
\end{thebibliography}

\end{document}